\documentclass[11pt]{article}
\usepackage[margin=1in]{geometry}
\usepackage[T1]{fontenc}
\usepackage{newtxtext,newtxmath}
\usepackage{amsmath,mathtools}
\usepackage{booktabs,longtable,array,tabularx,multirow}
\usepackage{enumitem}
\usepackage{graphicx}
\graphicspath{{figures/}}
\usepackage{xcolor}
\usepackage{tikz}
\usetikzlibrary{arrows.meta,positioning,fit,backgrounds,shapes.geometric,calc,decorations.pathreplacing,matrix}
\usepackage{listings}
\usepackage{algorithm}
\usepackage{algpseudocode}
\usepackage{hyperref}
\usepackage{cleveref}
\usepackage{microtype}
\usepackage{caption}
\usepackage{placeins}
\usepackage{subcaption}
\usepackage{csquotes}
\usepackage[backend=biber,style=numeric,sorting=none,maxbibnames=99]{biblatex}
\AtBeginBibliography{\footnotesize\sloppy}

\definecolor{deepblue}{RGB}{36,84,140}
\definecolor{midblue}{RGB}{89,139,190}
\definecolor{lightblue}{RGB}{224,236,248}
\definecolor{softgray}{RGB}{245,246,248}
\definecolor{darkgray}{RGB}{70,76,82}
\definecolor{accentred}{RGB}{176,53,57}
\definecolor{teal}{RGB}{40,120,120}

\hypersetup{colorlinks=true,linkcolor=deepblue,citecolor=deepblue,urlcolor=deepblue,
  pdftitle={Testing Agentic Service Systems: An Agentic Living Lab Methodology Across the Physical-Digital Continuum},
  pdfauthor={Zhongjie Wang, Xiang He, Xiao Wang, Mingyi Liu, Shuang Yu, Hanchuan Xu}}
\setlist[itemize]{leftmargin=2em,itemsep=0.15em,topsep=0.2em}
\setlist[enumerate]{leftmargin=2em,itemsep=0.15em,topsep=0.2em}

\newcolumntype{Y}{>{\raggedright\arraybackslash}X}
\newcolumntype{Z}{>{\centering\arraybackslash}X}

\newcommand{\term}[1]{\textbf{#1}}
\newcommand{\code}[1]{\texttt{\detokenize{#1}}}

\newcommand{\worldkernel}{World Kernel}

\newcommand{\evidencebundle}{Evidence Bundle}

\newcommand{\actionintent}{ActionIntent}

\begin{document}

\title{Testing Agentic Service Systems:\\An Agentic Living Lab Methodology Across the Physical--Digital Continuum}
\author{Zhongjie Wang, Xiang He, Xiao Wang, Mingyi Liu, Shuang Yu, Hanchuan Xu\\
\small Faculty of Computing, Harbin Institute of Technology, China\\
\small \texttt{\{rainy,hexiang,wxlxq,liumy,yushuang,xhc\}@hit.edu.cn}}
\date{}
\maketitle

\begin{abstract}
Task success is insufficient for testing agents that act for principals, change persistent business state, and affect physical processes: a useful result may be unauthorized, leave obligations unresolved, or depend on the execution substrate. We propose an Agentic Living Lab methodology organized around three questions: what is the test unit, who and what participate, and under what reality is the test executed? The unit is a service episode containing delegated roles, obligations, world state, and evidence. The method separates agents under test from environment actors and distinguishes real humans, human traces, rule surrogates, and agent-based proxies. Realization Profiles, capability-declared Reality Substrate bindings, Grounding Profiles, and Semantic Delta disclosures make changes across software, enterprise systems, simulation, and physical settings explicit. A four-plane reference architecture separates experiment control, participants, world realization, and evidence evaluation, with cross-cutting authority and governance controls. Runtime semantics distinguish intent, normative admission, and factual reconciliation. Real stakeholder grounding, participant calibration or validation, and governed feedback are required for the Living Lab claim, but not synchronous human participation in every run. Released JSON Schemas, deterministic fixtures, and a limited reference runner make selected rules machine-checkable; they are not a complete platform or empirical validation. Constructed travel and industrial designs illustrate authorization and fulfillment failures, proxy sensitivity, and substrate changes. Agent quality, proxy validity, and cross-substrate effect stability remain empirical questions.
\end{abstract}


\section{Introduction}

Services Computing has traditionally provided abstractions for describing, discovering, composing, managing, and evolving distributed computational and business capabilities as services~\cite{papazoglou2003soc,papazoglou2008roadmap,huhns2005soc,singh2005book,sheng2014composition,bouguettaya2017manifesto}. Microservices and cloud-era service engineering changed deployment scale and operational structure without removing the need to reason about composition, interaction, and service obligations~\cite{dragoni2017microservices}. The emergence of autonomous software agents changes a central operational assumption: the software actor itself may now interpret goals, select means, negotiate with counterparties, invoke heterogeneous services, revise plans, and act for an identifiable principal.

This shift changes not only how services are implemented but also how they should be \emph{tested}. A travel agent may reach the correct destination while purchasing without the required approval. A production-recovery agent may meet a delivery objective while violating a machine constraint or leaving a supplier obligation unresolved. A run may complete while a remote effect remains uncertain. The same agent may look strong in a software-only world and weak once enterprise timing, FMU dynamics, or human intervention enter the episode. Recent relationship-centered Services Computing work likewise distinguishes task completion from authorized action, fulfilled obligations, and realized value~\cite{wang2026sc30}. The present paper asks the methodological question that follows:

\begin{quote}
\textbf{How should agentic service systems be tested when software actors act on behalf of principals, participate in persistent service relationships, interact with humans and institutions, and may ultimately affect physical reality?}
\end{quote}

We organize the answer around three testing questions.

\begin{description}[leftmargin=1.7cm,style=nextline]
\item[TQ1 --- What is tested?] The unit is a \term{service episode}, not merely an Agent--Task pair. A test may need to preserve principals, role/carrier bindings, delegation, collaboration obligations, persistent world state, and evidence over time.
\item[TQ2 --- Who and what participate?] The \term{system under test} (SUT) is separated from test-harness actors. The latter include environment agents, conventional services, and human roles that may be enacted by real participants, human-trace replay, rule-based surrogates, or dynamically provisioned agent-based human proxies.
\item[TQ3 --- Under what reality is the test executed?] Participant and world objects may be realized digitally, through twins or replay, in shadow mode, through FMU/HIL components, or in a physical field setting. Their grounding and any substrate-induced semantic changes must be reported explicitly.
\end{description}

The human factor is essential. Real service systems contain people who delegate authority, approve or reject proposals, revoke permissions, take over execution, provide observations, adjudicate disputes, and accept closure. Those decisions are not peripheral annotations; they are episode events that can change what an agent is allowed to do and whether an outcome counts as fulfilled. Yet repeated experiments cannot recruit a large human panel for every Run. The methodology therefore treats \emph{human-role realization} as a configurable part of the test environment. A traveler, approver, operator, reviewer, or other human role can be enacted by a real participant when the research question requires it, or by a trace/rule/agent surrogate in high-throughput runs. The surrogate is a test-harness actor, not the agent under test unless the Study explicitly says otherwise.

This distinction is already emerging in interactive agent evaluation. $\tau$-bench simulates users with language models in tool-agent-user interactions, while $\tau^2$-Bench gives both the agent and user the ability to act in a shared dynamic environment~\cite{yao2025taubench,barres2025tau2}. At the same time, work on generative human proxies, SimulatorArena, and grounded user simulation shows that simulated users can differ systematically from real humans and can change conclusions about assistant quality~\cite{park2023generativeagents,dou2025simulatorarena,zhu2026realusersim}. We therefore do not treat ``an LLM user'' as a transparent substitute for a person. Participant realization and proxy sensitivity become part of the experimental contract.

We use the term \term{Agentic Living Lab} for the reference infrastructure that executes this testing methodology. The Lab is not a claim that every experiment must contain real humans or physical assets synchronously. Rather, a complete Living Lab must have technical interfaces through which real stakeholders can ground goals and constraints, participate in co-creation, calibrate or challenge proxies, authorize or review consequential decisions, and feed incidents back into future research assets. High-throughput Runs may still instantiate many stakeholder roles as controlled proxies. An environment without these stakeholder-grounding and feedback interfaces is \emph{Living-Lab-ready}, not a complete Living Lab. In particular,
\begin{equation}
ProtocolConformance \not\Rightarrow LivingLabStatus.
\label{eq:labeligibility}
\end{equation}
Protocol conformance is a property of the testing interface; Living-Lab status additionally requires the declared socio-technical bindings to real stakeholders, human grounding, and governed feedback.

This repositions the paper. It is not primarily a systems paper claiming a finished platform, and it is not another generic Agent benchmark. It is a \term{testing-method and reference-infrastructure report} for agentic service systems. The architecture exists to make the method executable and repeatable; the scientific claims are about what a valid test must control, observe, disclose, and distinguish.

The paper makes four top-level contributions:
\begin{enumerate}
  \item \textbf{Episode-centric testing and a multi-judgment oracle.} We define the service episode as the first-class test unit and distinguish Run completion, task success, service fulfillment, stakeholder value, authority, continuity, evidence, and human work rather than collapsing them into a single score.
  \item \textbf{Participant realization for human-in-the-loop testing.} We separate Agents Under Test from environment agents and human-role surrogates, define Real Human / Human Trace Replay / Rule Surrogate / Agent Proxy realizations, and require proxy-grounding and sensitivity disclosure when simulated human behavior materially affects a conclusion.
  \item \textbf{Reality-aware test configuration.} Object-level Realization Profiles, versioned Reality Bindings with capability-declared substrate adapters, multi-dimensional Grounding Profiles, normative-versus-factual authority, and Semantic Delta make changes in software/twin/shadow/FMU/HIL/physical realizations explicit. These are disclosure and control objects, not a claim of sim-to-real equivalence.
  \item \textbf{A reference test harness and implementation-independent conformance protocol.} We specify Experiment Control, Participant \& Agent, World \& Reality, and Evidence \& Oracle planes; a fail-closed runtime kernel; version-pinned Run Manifests and Evidence Records; deterministic conformance fixtures; and a small ancillary reference runner for the currently released subset.
\end{enumerate}

The paper deliberately avoids component-priority claims. Delegation, social commitments, user simulators, Living Labs, digital twins, policy enforcement, FMI/FMU, provenance, event sourcing, and testbeds all have established precedents. The increment is the way they are bound into one testing contract for a versioned service episode. The internal cognitive architecture of the SUT is also out of scope: an AUT may be an LLM agent, a symbolic planner, a workflow, a classical MAS component, or a hybrid. Normative semantics stop at Observation, ActionIntent, AdmissionDecision, effect/reconciliation, and Evidence; private chain-of-thought is neither required nor treated as authoritative evidence.

The travel and industrial cases are constructed experiment designs rather than empirical validation. They show how the methodology diagnoses relational failures, varies human-role realizations, separates proxy effects from AUT effects, and separates substrate sensitivity from agent improvement. Comparative agent-quality claims, human-proxy validity, and cross-substrate effect stability remain empirical questions.
\begin{figure}[t]
\centering
\includegraphics[width=0.98\linewidth]{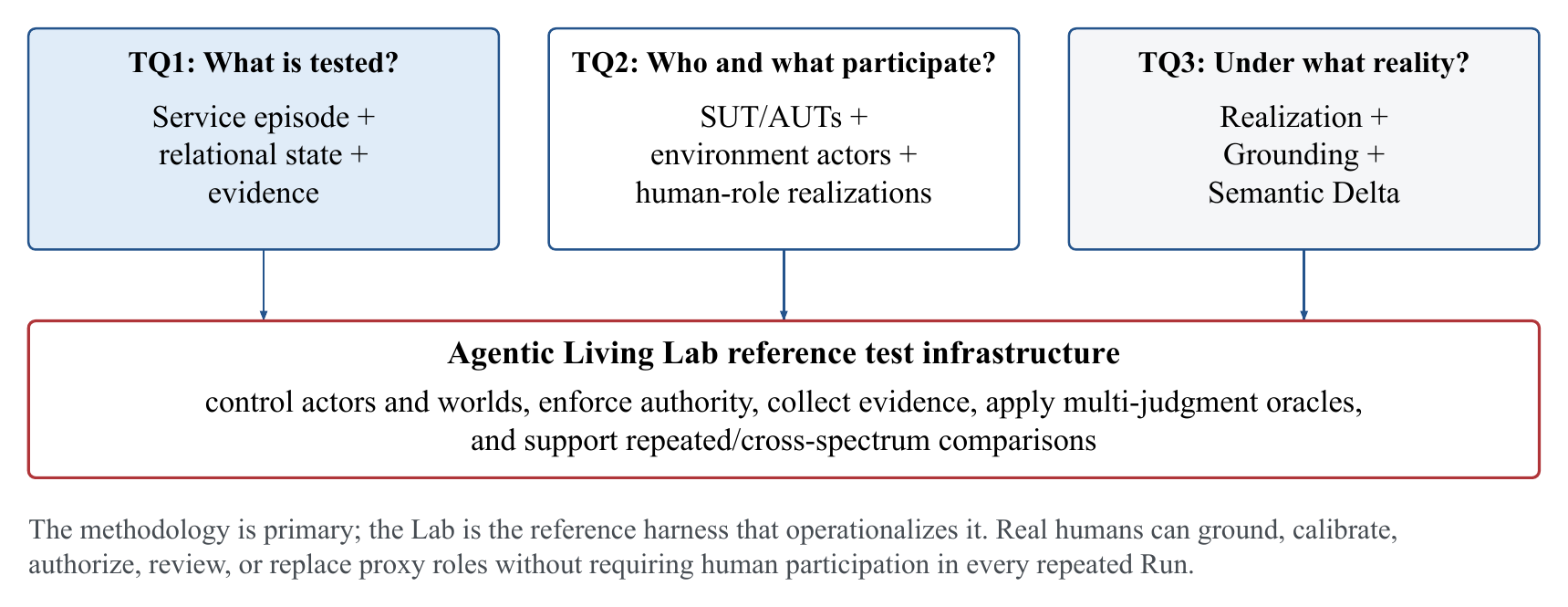}
\caption{Testing perspective of the paper. The contribution is a method for testing agentic service systems; the Agentic Living Lab is the reference infrastructure that makes the method executable across human and physical--digital realizations.}
\label{fig:testquestions}
\end{figure}
\section{Testing Model for Agentic Service Systems}

\subsection{Service episode as the first-class test unit}

Commitment-based service research has long shown that flexible service interaction cannot be reduced to a sequence of interface calls~\cite{yolum2002commitment,singh2009commitment}. For testing purposes, a \term{service relationship} is a continuing arrangement among participants, their roles, applicable agreements, and institutional conditions. A \term{service episode} is a bounded enactment directed toward an outcome under one or more such relationships. One corporate-travel relationship may support many trips; one trip may involve several provider relationships.

The test harness does not require a new ontology for these objects. It requires a minimum episode record that remains stable enough for comparison: an episode identifier; principals, roles, and active role--carrier bindings; relevant relationship identifiers; the initial world snapshot; applicable delegation, collaboration-contract, and policy versions; an authoritative sequence of participant, environment, action, observation, reconciliation, and review events; and retained evidence/provenance references. Observations may be incomplete, delayed, or contradictory; the record preserves that uncertainty rather than replacing it with an agent-generated narrative.

The methodological shift is therefore
\begin{equation}
Agent\times Task\rightarrow Score
\quad\Longrightarrow\quad
Episode\times World\times Relationships\times Actors\times Evidence\rightarrow Judgments.
\label{eq:unitshift}
\end{equation}
The right-hand side does not claim that every test must be long-running or cross-organizational. It says that when authority, counterparty obligations, human decisions, or persistent effects matter, the test must retain enough episode state to judge them. In plain terms, the episode record pins down \emph{who acted for whom, under which relationship state, against which world version, with what evidence, and with what resulting judgment}.

\subsection{System under test, environment actors, and human-role realizations}

A key testing distinction is between the object being evaluated and the actors supplied by the test harness. We use \term{System Under Test (SUT)} for the agentic service system whose behavior is being judged. A SUT may contain one AUT or a set
\begin{equation}
\mathcal A_{SUT}=\{AUT_1,\ldots,AUT_n\}.
\label{eq:multiAUT}
\end{equation}
It may also contain conventional software and internal protocols owned by the tested system.

When multiple AUTs belong to the same SUT, their mutual messages, negotiations, shared-memory operations, and A2A exchanges are \emph{internal SUT behavior}. The harness may observe and retain these interactions for diagnostics or failure attribution, but it does not score them as environment-provided inputs. A Study can instead place one party outside the SUT---for example, a supplier agent representing an independent enterprise---in which case the same protocol exchange crosses the SUT/environment boundary and becomes an externally provisioned interaction. The Run Manifest therefore freezes actor membership, not merely actor names.

Actors outside the SUT are divided into environment/counterparty actors and human-role realizations. Software actors that realize human roles remain harness actors unless the Study explicitly declares them part of the tested system. Real human participants remain distinct from software proxies.

A human role $r_h$ is instantiated through a \term{Participant Realization}:
\begin{equation}
\rho_H(r_h)\in\{RealHuman, HumanTraceReplay, RuleSurrogate, AgentProxy\}.
\label{eq:humanrealization}
\end{equation}
These modes are not a realism ladder. They answer how a required human role is enacted in a particular Run.

\begin{table}[t]
\centering
\caption{Human-role realizations in an agentic-service test. The role semantics remain part of the episode even when a surrogate enacts the role.}
\label{tab:humanrealization}
\scriptsize
\begin{tabularx}{\textwidth}{p{3.0cm}p{4.0cm}Y}
\toprule
Realization & Typical use & Test obligation\\
\midrule
Real Human & co-creation, approval, takeover, acceptance, dispute, high-grounding validation & consent/identity/role binding; protect participant data; record the decision as an episode event\\
Human Trace Replay & repeat a historically observed decision or interaction sequence & preserve provenance, context, and limits of reuse; do not infer unseen human responses\\
Rule Surrogate & deterministic or stochastic scripted user/approver/operator behavior & freeze rule/version/seed; report its behavioral scope\\
Agent Proxy & dynamically provisioned software agent acting as traveler, approver, operator, reviewer, or other human role & freeze model/prompt/policy/memory/seed as applicable; disclose calibration; separate proxy behavior from AUT behavior\\
\bottomrule
\end{tabularx}
\end{table}

Human decisions that change authority or service state remain first-class episode events: delegation, approval, rejection, revocation, preference amendment, takeover, dispute, review, and closure acceptance. Replacing a real person with a proxy changes who enacts the role, not the fact that these decisions must be represented and evidenced.

\subsection{Multi-judgment test oracle: capability is not fulfillment}

A central implication is that several judgments often collapsed into ``success'' do not entail one another:
\begin{equation}
RunCompletion\;\not\Rightarrow\;TaskSuccess\;\not\Rightarrow\;ServiceFulfillment\;\not\Rightarrow\;StakeholderValue.
\label{eq:gap}
\end{equation}
A completed run may contain a rejected or uncertain action. A task goal may be achieved through an unauthorized purchase. A technically successful recovery may leave a counterparty obligation unresolved. A service can be fulfilled while stakeholder value is negative because supervision or recovery cost dominates.

Accordingly, a test result is reported as a non-aggregated vector rather than collapsed into a single scalar:
\begin{equation}
\mathbf J=\langle O,A,C,R,E,H,V\rangle,
\label{eq:reportvector}
\end{equation}
where $O$ captures outcome/obligation state, $A$ authority and constraint compliance, $C$ collaboration/commitment state, $R$ continuity and recovery, $E$ evidence and contestability, $H$ human work/supervision, and $V$ stakeholder value. Mandatory violations remain visible and are not hidden in a weighted average. Section~\ref{sec:metrics} gives operational definitions for selected components rather than treating the vector as a purely descriptive checklist.

The human-work component is now explicit. For real-human conditions it may include approval latency, intervention rate, takeover burden, review time, or disagreement. For human proxies it includes proxy-induced interaction patterns and the sensitivity of conclusions to the chosen proxy realization. A proxy that always cooperates can make an AUT look artificially reliable; an adversarial or over-literal proxy can make it look artificially weak. A Study that relies materially on simulated human behavior must therefore either justify the proxy against human evidence or report a proxy-sensitivity comparison.

\subsection{R1--R4 as observable testing boundaries}

The protocol uses four operational boundaries that are consistent with commitment/delegation research and with recent relationship-centered Services Computing discussions~\cite{yolum2002commitment,singh2009commitment,wang2026sc30}. They are defined here as part of the testing method.

\begin{description}[leftmargin=1.2cm,style=nextline]
\item[R1: Admit before consequential execution.] A consequential action must pass current authority, prerequisite, policy, resource, and evidence checks before dispatch, and the decision must be recorded.
\item[R2: Create commitments through an authorized protocol.] A model-generated promise is not automatically a valid commitment. Debtor, creditor, activation conditions, required outcome/behavior, due conditions, evidence, and exit rules become effective only through the agreed authorization protocol.
\item[R3: Revoke prospective authority while preserving incurred obligations.] Revocation prevents new use of affected authority after its effective point but does not erase an already accepted purchase, promise, or uncertain remote action.
\item[R4: Close explicitly; distinguish fulfillment from termination.] Fulfilled closure requires accepted outcome evidence and discharged or validly transferred obligations. Failed, cancelled, breached, disputed, or unresolved episodes can terminate without being labeled fulfilled.
\end{description}

These rules become testable boundaries. A Study can inject stale credentials, delayed approval, user revocation, duplicate dispatch, missing evidence, partial provider acceptance, human takeover, or disputed observations and check whether the SUT and harness reach the expected state.

\subsection{Minimal semantic kernel}

\label{sec:minikernel}
Table~\ref{tab:vocab} defines the minimum vocabulary used by the testing protocol.

\begin{table}[t]
\centering
\caption{Minimal vocabulary required by the testing protocol.}
\label{tab:vocab}
\small
\begin{tabularx}{\textwidth}{p{3.2cm}Y}
\toprule
Term & Operational meaning in the test harness\\
\midrule
Principal & Person or organization whose interests and valid authority ground a delegation.\\
Service-agent role & Delegated digital role with bounded discretion for a principal.\\
Carrier & Software entity that realizes one or more roles; changing a carrier does not automatically transfer authority or obligations.\\
Service relationship & Continuing arrangement among participants, bindings, agreements, and institutional conditions.\\
Service episode & Bounded enactment under one or more relationships; the first-class test unit.\\
SUT / AUT & System or agent(s) whose behavior is being evaluated.\\
Environment actor & Counterparty/service actor supplied by the harness and not under test.\\
Human-role realization & Real human, trace replay, rule surrogate, or agent proxy enacting a human role.\\
ActionIntent & Proposed consequential action; never itself a world fact.\\
EpisodeEvent & Canonical participant, environment, action, observation, reconciliation, or review event.\\
Evidence Record & Canonical record of a consequential admission/effect/reconciliation boundary.\\
\bottomrule
\end{tabularx}
\end{table}

The \term{minimal runtime contract} consists of four implementation-independent interfaces:
\begin{center}
\begin{tabularx}{0.93\textwidth}{p{2.75cm}Y}
\toprule
Interface & Required behavior\\
\midrule
\texttt{Observe} & project the versioned world/evidence state into the information visible to a role; new live substrate information becomes a new observation event\\
\texttt{Admit} & evaluate all applicable mandatory guards over the pinned state and resolve them with the fail-closed algebra\\
\shortstack{\texttt{Execute}/\\\texttt{Reconcile}} & dispatch only admitted actions, observe effects through the bound substrate, and apply the versioned reconciliation policy when factual claims are uncertain or contested\\
\shortstack{\texttt{Evidence}\\\texttt{Commit}} & append canonical evidence for participant decisions, admission, dispatch/effect, reconciliation, and closure without rewriting earlier history\\
\bottomrule
\end{tabularx}
\end{center}

Each mandatory admission guard is evaluated in $\{PASS,FAIL,UNKNOWN\}$ and resolved fail-closed to
\begin{equation}
AdmissionDecision\in\{ALLOW,DENY,HOLD,ESCALATE\},
\label{eq:admission}
\end{equation}
with total priority
\begin{equation}
DENY\succ ESCALATE\succ HOLD\succ ALLOW.
\label{eq:priority}
\end{equation}
Hard failure dominates; an authority/institutional unknown with a declared review route yields ESCALATE; a remaining explicitly waitable unknown yields HOLD. ALLOW requires every applicable mandatory guard to pass; it permits proceeding to reservation, not immediate dispatch. Dispatch additionally requires successful reservation and current-version checks, as specified in Section~\ref{sec:runtime}. UNKNOWN is never silently treated as true.

Two safety properties are normative requirements. The released fixtures exercise selected decision-level consequences, not end-to-end enforcement by a running platform:
\begin{align}
Admission=DENY &\Rightarrow \neg Dispatch\land\neg Reservation, \label{eq:safetydeny}\\
MandatoryFact\in\{Contested,Unknown\} &\Rightarrow Closure\neq Fulfilled. \label{eq:safetyclosure}
\end{align}
HOLD records a wake condition, deadline, and expiry policy. ESCALATE identifies an authorized target and blocks the original action until fresh admission. The later runtime sections refine how participant decisions, resource races, remote effects, and reconciliation interact with this contract.

\section{Positioning Against Existing Evaluation and Experimental Paradigms}

\subsection{Stateful agent benchmarks and simulated users}

Interactive agent benchmarks already go well beyond static question answering. AgentBench, AgentBoard, and OSWorld expose multi-step interaction and trajectory behavior~\cite{liu2024agentbench,ma2024agentboard,xie2024osworld}. CRMArena, TheAgentCompany, and STATE-Bench are explicitly stateful enterprise environments with interconnected records, policies/procedures, task-local databases, and state-changing actions~\cite{huang2025crmarena,xu2024agentcompany,microsoft2026statebench}. The distinction made by this paper is therefore \emph{not} that current benchmarks are stateless or short-lived.

Interactive user simulation is also established. $\tau$-bench evaluates tool-using agents while a language-model user participates in domain-constrained conversations; $\tau^2$-Bench goes further by allowing both the tested agent and user to act on a shared dynamic environment~\cite{yao2025taubench,barres2025tau2}. These systems are particularly close to the Participant \& Agent Plane proposed here. They demonstrate that a ``user'' can be an active environment actor rather than a passive prompt source.

The testing gap addressed here is narrower: a versioned episode protocol in which principal/role binding, delegation, cross-party commitments, human-role realization, factual status, closure, evidence, and reality/grounding declarations are portable and conformance-testable across environments. A sufficiently extensible enterprise benchmark could implement this protocol and become a highly frozen mode of an Agentic Living Lab. Adding a few database tables is not excluded; it is simply insufficient unless the corresponding transition, evidence, and reporting semantics are enforced.

The use of human proxies introduces its own validity problem. Generative-agent work shows that LLM-based actors can enact plausible human-like behavior~\cite{park2023generativeagents}, while SimulatorArena and RealUserSim show that simulator design and grounding can materially affect how closely simulated-user evaluations align with real-human judgments~\cite{dou2025simulatorarena,zhu2026realusersim}. The methodological response in this paper is not to reject proxies but to make their realization, calibration, and sensitivity part of the test contract.

\subsection{Living Labs, testbeds/TEFs, and real-world MAS}

Living Lab research contributes real or socially meaningful problems, stakeholder co-creation, longitudinal operation, and iterative feedback. The European Network of Living Labs describes Living Labs as user-centered open-innovation ecosystems integrating research and innovation in real-life communities and settings~\cite{enoll2026}. This social-technical grounding remains necessary for the present use of the term. A synthetic harness with no stakeholder grounding or governed feedback is Living-Lab-ready, not automatically a complete Living Lab.

Open testbeds and TEFs contribute a different capability. SmartSantander demonstrated city-scale experimentation over heterogeneous real infrastructure~\cite{sanchez2014smartsantander}. The European Commission defines AI Testing and Experimentation Facilities as combinations of physical and virtual facilities for testing AI software, hardware, and robots in realistic environments~\cite{euteF2026}. Neither field experimentation nor physical--virtual hybrid facilities are novelty claims of this paper.

MAS has also entered real Living Lab and industrial settings. PowerMatching City coordinated real households and energy devices through agent-based mechanisms~\cite{bliek2010powermatching}; SmartFactory-KL-related work deployed holonic multi-agent control in an Industry 4.0 demonstrator~\cite{bernhard2023holonic}. Again, the increment is not ``agents in the real world.'' It is a test contract that can keep SUT identity, environment actors, human-role realizations, authority, obligations, world effects, and evidence aligned as the realization changes.

\subsection{Digital twins, AgentOps, normative systems, and reproducibility objects}

The boundary to digital-twin research is increasingly close. The Digital Twin Consortium's 2026 Digital Twin System Framework includes autonomous agents as actuators and emphasizes authority, evidence, simulation gates, and governed action~\cite{dtc2026framework}; digital-thread research emphasizes lifecycle linkage, provenance, interoperability, and traceable engineering decisions~\cite{wu2025digitalthread}. AgentOps focuses on observability and lifecycle management for agentic systems~\cite{dong2024agentops}. W3C PROV and SIMPROV provide established provenance mechanisms~\cite{moreau2013prov,ruscheinski2025simprov}; RO-Crate and CWL package research artifacts and portable workflows~\cite{soilandreyes2022rocrate,crusoe2022cwl}. Runtime contract-compliance research monitors obligations in service settings~\cite{lomuscio2011monitoring}.

Normative multi-agent systems and usage-control models already treat institutional conditions, permissions, obligations, and ongoing control as first-class objects~\cite{boella2006normative,park2004ucon}. Artifact-centric business-process models give long-running business entities explicit lifecycle semantics~\cite{hull2011gsm}. These are direct antecedents. The present contribution is synthesis plus operational binding: the same version-pinned service episode carries the tested actors, delegation/commitment state, participant realizations, factual reconciliation, reality/grounding disclosure, and evidence needed for a test verdict.

\subsection{Compact positioning}

\begin{table}[t]
\centering
\caption{Positioning by primary test object and first-class semantics. ``Strong'' means a central organizing concern, not that other paradigms cannot implement the capability.}
\label{tab:positioning}
\scriptsize
\begin{tabularx}{\textwidth}{p{2.6cm}p{2.8cm}ZZZZ}
\toprule
Paradigm & Primary unit & Persistent world & Human actor realization & Org./contracts & Freeze/replay / multi-substrate\\
\midrule
Classical service test/workflow & Service / process & Partial & Usually external & Partial & Strong / limited\\
Stateful agent benchmark & Agent/SUT $\times$ task & Strong within environment & Increasingly simulated/active & Policy-dependent & Strong / mainly digital\\
Agent runtime/platform & Agent/application & External & External & Partial & Not central / external\\
MAS simulation & Agent society & Strong (simulated) & Synthetic agents & Simulated rules & Strong / digital\\
Digital Twin System & Asset/system & Strong & Operator dependent & Authority increasing & Digital thread / strong\\
AI TEF/testbed & Technology under test & Real environment & Can be real & Scenario-dependent & Facility-dependent / strong\\
Traditional Living Lab & Innovation ecosystem & Strong & Real stakeholders central & Socio-technical governance & Not central / strong\\
\textbf{Agentic Living Lab method} & \textbf{Service episode / SUT} & \textbf{First-class} & \textbf{Declared participant mode} & \textbf{Delegation + collaboration} & \textbf{First-class / object-level}\\
\bottomrule
\end{tabularx}
\end{table}

The paper's claim is therefore methodological rather than exclusivist. Existing benchmarks, workflow engines, agent platforms, digital twins, or testbeds can host the method if they preserve the required episode semantics and pass the corresponding conformance conditions. The Agentic Living Lab names the broader socio-technical testing infrastructure in which those test modes can be combined with stakeholder grounding and feedback.

\section{Agentic Living Lab Testing Methodology: Participants, Grounding, and Reality}

\subsection{Definition and three forms}

We define an \term{Agentic Living Lab} as a long-lived socio-technical testing infrastructure in which people, organizations, agents, services, and resources participate in governed service episodes; SUT behavior is exercised through explicit actor, authority, world, and oracle configurations; experimental objects can be realized on software, twin, shadow, HIL, or physical substrates; and real stakeholder feedback can be reviewed and promoted into future test assets. The use of \emph{Living Lab} follows the established emphasis on stakeholder involvement, co-creation, and real-life grounding~\cite{enoll2026,leminen2012livinglabs,hossain2019livinglabs}.

Implementing the technical conformance protocol alone does \emph{not} make an environment an Agentic Living Lab. A complete Lab must provide three socio-technical bindings: (i) real-stakeholder participation in problem/scenario or acceptance-criterion definition, (ii) real-human or human-grounded calibration/validation of participant roles when human behavior matters, and (iii) a governed path by which real episodes/incidents can become later experimental assets. Without these conditions the environment is \term{Living-Lab-ready}, regardless of how sophisticated its simulation or hardware is.

We therefore distinguish three forms:
\begin{description}[leftmargin=0.6cm,style=nextline]
\item[Living-Lab-ready experimental environment.] The episode protocol, world state, test control, evidence, and replay are available, but stakeholder co-creation/grounding or governed real feedback is absent.
\item[Digitally grounded Agentic Living Lab.] Runs may be entirely digital, but real stakeholders, operational rules/data, calibrated human behavior, or governed operational feedback ground the test program.
\item[Physically situated Agentic Living Lab.] Physical participants, assets, environments, or consequential physical actions are direct components of the Study.
\end{description}
The forms are not ordered by maturity. A digitally grounded Lab can be the appropriate final form for a digital-native service domain.

A full Living-Lab claim leaves a technical footprint. Table~\ref{tab:livingfootprint} summarizes the minimum interfaces used by this paper.

\begin{table}[t]
\centering
\caption{Minimum technical footprint for a complete Agentic Living Lab.}
\label{tab:livingfootprint}
\scriptsize
\begin{tabularx}{\textwidth}{p{3.2cm}Yp{3.8cm}}
\toprule
Footprint & Required meaning & Example record/interface\\
\midrule
Stakeholder participation & who participates, in which role, under what consent/data conditions & participant/consent record linked to Study\\
Co-creation/change input & stakeholder-originated goals, failure criteria, rules, or redesign requests are attributable & change proposal with provenance and decision\\
Human decision interface & delegation, approval, revocation, takeover, review, dispute, and closure can be performed or inspected & Real Human Bridge / review console / signed decision\\
Proxy calibration/challenge & surrogate human behavior has a documented basis and can be challenged & calibration asset, proxy version, disagreement record\\
Feedback assetization & incidents become research assets only through governed review & incident-to-Scenario/Rule/GoldenRun promotion record\\
\bottomrule
\end{tabularx}
\end{table}

\subsection{Human participation as a configurable test dimension}

Real people are first-class service participants, but they do not need to be physically or synchronously present in every repeated Run. The Lab therefore separates the \emph{semantic human role} from its current realization. The same traveler role, for example, can be enacted by a real person during co-creation and acceptance, by a historical trace in a replay study, by a deterministic rule surrogate for unit-like conformance tests, or by an agent proxy for large-scale interactive evaluation.
\begin{figure}[t]
\centering
\includegraphics[width=0.80\linewidth]{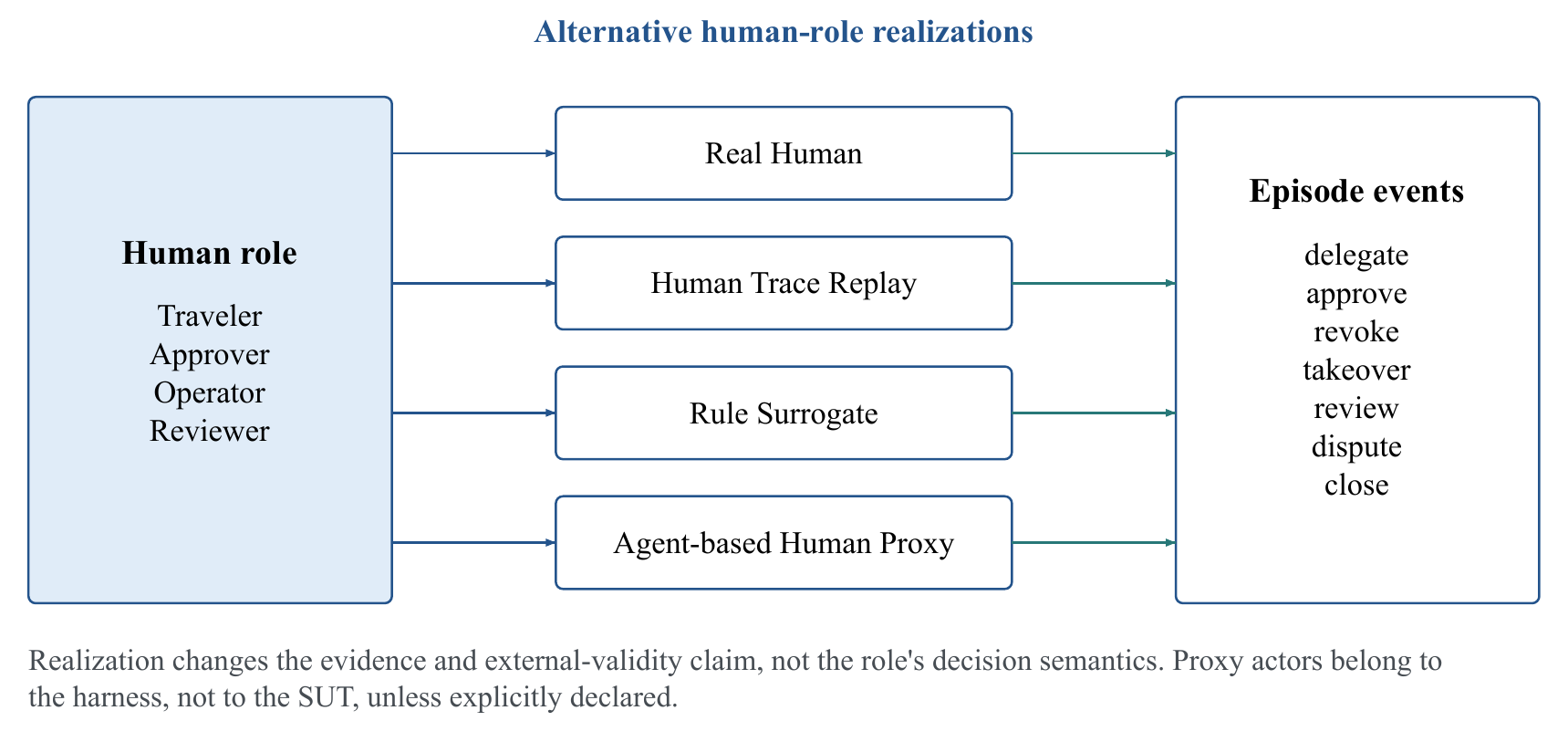}
\caption{Human-role realization. A Study can substitute execution carriers for a human role while keeping delegation/approval/review semantics explicit. The substitution itself is recorded and grounded.}
\label{fig:humanroles}
\end{figure}

For an Agent Proxy, the Run Manifest freezes the model/runtime, prompt or policy, memory state if used, random seed, tool access, and any behavior/calibration profile that can affect interaction. A Lab may dynamically provision many such proxies to create diverse user, approver, operator, or reviewer populations. This scaling mechanism is useful precisely because those proxies are \emph{not} the measured AUT.

Any human dataset used to calibrate a proxy or compare proxy behavior with real participants is itself a versioned experimental asset. The Participant Realization Profile therefore records a calibration-asset reference/hash, population and sampling description, behavioral dimensions used, calibration procedure, consent/use scope, access policy, release status, and privacy constraints. If the calibration data can be released under a compatible policy, the artifact should accompany the study. If privacy or institutional constraints prevent release, a public de-identified/shadow artifact may be supplied as a reproducibility aid but must not be represented as semantically equivalent to the private calibration data. In that case, proxy-validity and external-validity claims are explicitly bounded to the declared private calibration context.

The substitution is also a threat to validity. If the AUT and all human proxies use the same model family, they may share knowledge, style, or systematic biases. If a proxy is overly cooperative, the AUT can look artificially competent; if it over-amplifies behavioral directives, the AUT can look artificially brittle. Hence a Study whose conclusion depends materially on proxy behavior should conduct \term{proxy-sensitivity analysis}: rerun a matched condition with at least one alternative proxy realization or justify the proxy against human traces/participants, and report whether the direction/ranking of the SUT result changes.

\subsection{Realization and grounding}

A whole test is rarely located at one point on a physical--digital continuum. An industrial Study may contain a real planner, a software procurement AUT, a real ERP in read-only shadow mode, an MES twin, an emulated WMS, an FMU for a machine, and one HIL robot. A travel Study may be fully software-native while its goals and acceptance criteria are grounded by real users.

The method therefore keeps two first-class configuration \emph{disclosures} here. A \term{Realization Profile} records where each relevant world/service/asset or participant role is instantiated, using categories such as Digital, Emulated, Twin, Shadow, FMU/HIL, Physical, Real Human, Trace Replay, Rule Surrogate, or Agent Proxy. A \term{Grounding Profile} records what real-world evidence supports the interpretation of those realizations. The execution-level \term{Reality Binding} introduced below operationalizes the world/service/asset part of the Realization Profile; it does not replace either disclosure. ``Physicality'' is a derived description of a Realization Profile, not a maturity score. Domain-specific model fidelity may be reported when the Study has a defensible fidelity measure, but the methodology does not define a universal fidelity axis. ``Completeness'' is reserved only for the eligibility conditions that distinguish a complete Living Lab from a Living-Lab-ready environment.

\begin{table}[t]
\centering
\caption{Grounding dimensions. Categories are descriptive declarations rather than a universal maturity ladder.}
\label{tab:grounding}
\scriptsize
\begin{tabularx}{\textwidth}{p{2.8cm}Y}
\toprule
Dimension & Example declarations\\
\midrule
Stakeholder grounding & synthetic persona; rule surrogate; agent proxy; human-trace-calibrated proxy; real-human shadow; active real participant\\
Data grounding & synthetic; de-identified historical; calibrated historical; live shadow; live operational\\
Process grounding & toy; expert-specified; document-derived; production-policy-derived; live process\\
Parameter grounding & assumed; expert-estimated; historically calibrated; field-calibrated\\
Outcome grounding & simulated; twin-observed; shadow-compared; physically observed; institutionally accepted\\
\bottomrule
\end{tabularx}
\end{table}

These declarations are intentionally not collapsed into one scalar. ``Tested on real hardware,'' ``tested with real users,'' and ``validated against real business outcomes'' are different claims. Where a Study uses a domain-specific fidelity metric (for example, an FMU calibration error), that metric is reported alongside the Realization/Grounding Profiles rather than treated as a universal Lab maturity level.

\subsection{The physical--digital continuum and Reality Substrate}

The two ends of the continuum are ideal types. At the physical end, relevant consequences occur primarily through people, assets, and processes in real settings. At the software-native end, experimental entities and consequences are instantiated digitally. Most useful studies are mixed.
\begin{figure}[t]
\centering
\includegraphics[width=0.98\linewidth]{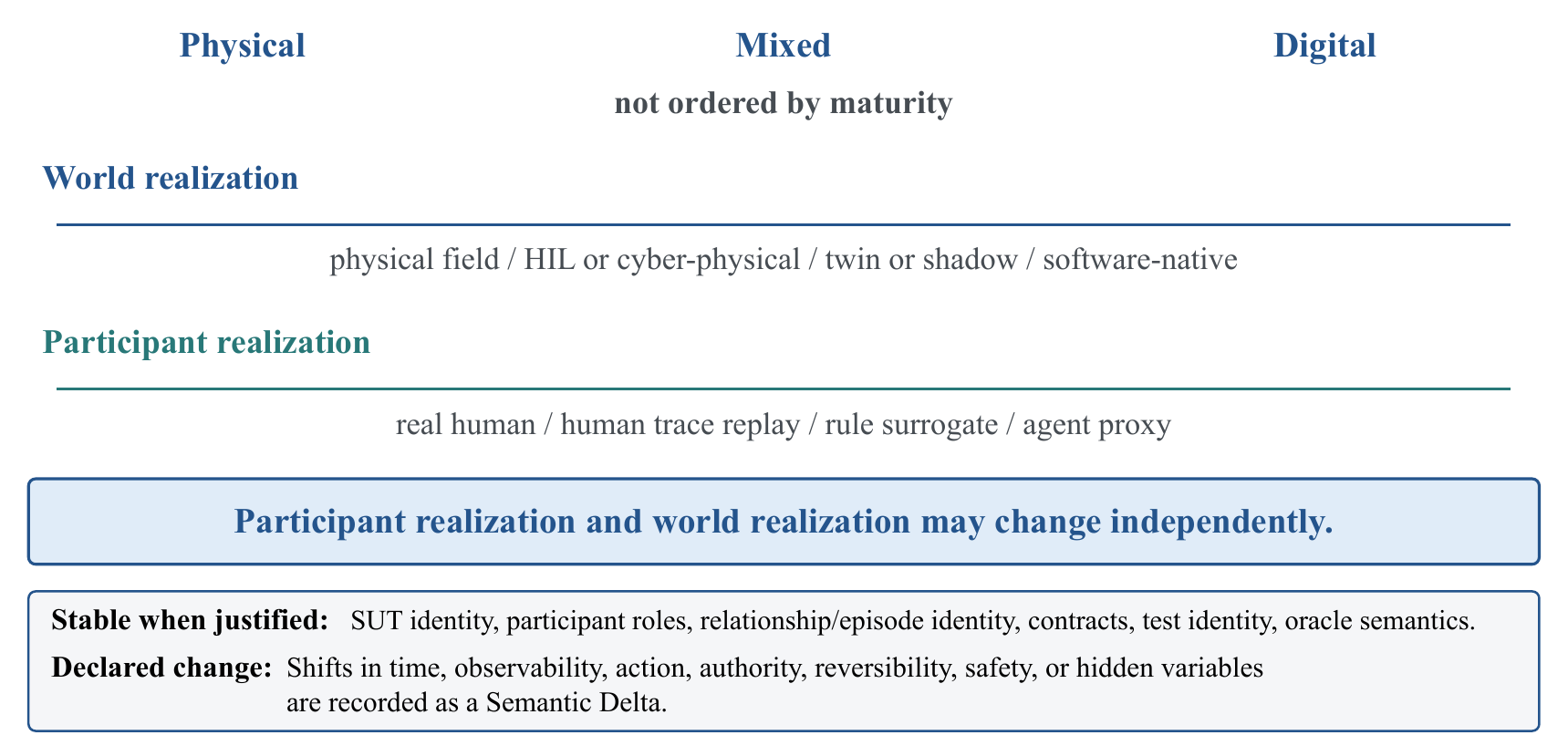}
\caption{Physicality is a test configuration variable, not a maturity ladder. Participant and world realizations can change independently.}
\label{fig:continuum}
\end{figure}

A \term{Reality Substrate} is an adapter whose declared capabilities may include observe, act, synchronize, logical-time advance, snapshot, reset, fork, and compensation. Not every substrate implements every capability. A software world can often snapshot/reset/fork; a shadow connection may be observe-only; HIL permits bounded action with partial reset; physical reality generally cannot be forked or rolled back.

\begin{table}[H]
\centering
\caption{Typical substrate capabilities; actual adapter capabilities are declared per Study.}
\label{tab:substratecaps}
\scriptsize
\begin{tabularx}{\textwidth}{p{2.2cm}ZZZZZ}
\toprule
Capability & Software & Enterprise Twin & FMU & HIL & Physical field\\
\midrule
Observe & yes & yes & yes & yes & yes\\
Act & yes & yes & yes & bounded & bounded\\
Logical-time advance & yes & partial & yes & limited & no\\
Snapshot/reset & strong & strong/partial & strong & partial & usually no\\
Fork & strong & strong & strong & digital side only & no\\
Exact replay & strong & conditional & conditional & weak & no\\
Physical consequence & no & no & no & partial & yes\\
Compensation & digital inverse/transaction & business compensation & model action & physical/business & physical/business\\
\bottomrule
\end{tabularx}
\end{table}

\subsection{A minimal Reality Substrate contract}

The substrate boundary is not only a taxonomy. It is a versioned execution protocol that allows stable experiment objects and runtime judgments to be retained while selected world objects move from software to enterprise shadow/twin, FMU, HIL, or physical realizations. Before planning or launching a Run, every substrate adapter exposes a \term{Capability Descriptor}. The descriptor identifies the adapter and version, declares supported operations and their constraints, and records relevant time, state, reversibility, side-effect, authority, and safety semantics. A required capability that is absent, inconsistent with the binding, or unavailable at launch causes planning or launch to fail closed; an implementation must not silently emulate that capability and preserve the original label.

Table~\ref{tab:substratecontract} defines the minimum implementation-independent contract. Operation names are canonical semantics rather than a prescribed programming language or transport. Only \code{describeCapabilities}, \code{bind}, and \code{observe} are universally required. Other operations are conditional on the descriptor and on the Study's authority and safety policies.

\begin{table}[H]
\centering
\caption{Minimum Reality Substrate contract. Optional operations are usable only when declared by the Capability Descriptor.}
\label{tab:substratecontract}
\scriptsize
\begin{tabularx}{\textwidth}{p{3.0cm}p{2.1cm}Y}
\toprule
Canonical operation & Availability & Contract meaning\\
\midrule
\code{describeCapabilities} & mandatory & Return adapter identity/version, supported operations, constraints, time/state semantics, effect-receipt semantics, and authority/safety requirements.\\
\code{bind(bindingSpec)} & mandatory & Bind one declared world/service/asset object to versioned endpoint/artifact, input/output or channel mappings, policies, and Semantic Delta reference.\\
\code{observe(query)} & mandatory & Return provenance-bearing observations under the binding's visibility and time semantics; an observation is evidence input, not automatically an accepted fact.\\
\code{dispatch}\newline\code{(admittedCommand)} & conditional on act & Accept only an admitted command carrying current authority/reservation/binding references; distinguish received, dispatched, effect-observed, unknown-effect, and contested states.\\
\code{synchronize}/\code{advance} & optional & Synchronize clocks or request logical/substrate-time progress under declared stepping and lateness constraints.\\
\code{snapshot}/\code{reset}/\code{fork} & optional & Capture, restore, or branch state only within the declared state boundary; physical rollback must not be inferred from a digital snapshot.\\
\code{compensate} & optional & Request a declared digital, business, or physical compensating action; compensation is a new consequential action rather than erasure of history.\\
\bottomrule
\end{tabularx}
\end{table}

For each relevant non-participant object, the Run Manifest freezes a versioned \term{Reality Binding}:
\begin{equation}
\begin{aligned}
RealityBinding=\langle&object, bindingVersion, substrateType, adapter,\\
&capabilities, artifact/endpoint, I/O\ mapping, timeBinding,\\
&authorityPolicy, reconciliationPolicy, safetyPolicy, semanticDeltaRef\rangle.
\end{aligned}
\label{eq:realitybinding}
\end{equation}
Bindings are object-level rather than Run-level labels: the same Run may bind an ERP shadow, an MES twin, an FMU machine, an HIL robot, and a software counterparty. A binding cannot be silently replaced during a Run. Rebinding increments the binding version, creates evidence, and triggers the corresponding Semantic Delta and comparability checks.

The adapter does not decide that a dispatch receipt or raw sensor value is an authoritative business fact. It reports observations and effect states; the versioned Reconciliation Policy in the runtime determines accepted, contested, or unknown fact status and commits the supporting evidence. Likewise, participant carriers are not collapsed into this device/service interface. Real Human, Human Trace Replay, Rule Surrogate, and Agent Proxy remain Participant Realizations mediated by the Human/Participant Bridge, because identity, consent, delegation, revocation, and decision rights are part of their contract.

The resulting separation is therefore
\begin{equation}
\begin{aligned}
AgenticLivingLab ={}& StableExperimentSemantics\\
&+ CapabilityDeclaredRealityAdapters\\
&+ VersionedObjectLevelBindings\\
&+ ExplicitSemanticDelta.
\end{aligned}
\label{eq:substrateseparation}
\end{equation}
``Stable'' here means that episode, participant, authority, evidence, and oracle objects retain their interface and identity across a migration. It does not assert that two substrates have identical timing, observability, reversibility, safety, hidden state, or causal meaning. Run planning and admissible operations are rebound to the declared capabilities, and cross-substrate interpretation remains governed by Semantic Delta.

\subsection{Semantic Delta and comparability}

Moving from substrate $R_i$ to $R_j$ does not imply that the test remains semantically identical. The method therefore requires a \term{Semantic Delta} disclosure covering changes in time semantics, observation semantics, action binding, authority, reversibility, safety/failure modes, and hidden variables. Each item is classified for the target hypothesis as \emph{comparison-preserving}, \emph{comparison-with-caveat}, or \emph{comparison-breaking}. The classification is made by the Study designer before interpreting the comparison and must cite the concrete changed artifact/policy; it is a disclosure and comparability decision, not a general calculus for sim-to-real equivalence.

Table~\ref{tab:deltaworked} gives the full worked example used later in the industrial design. The AUT, business scenario, delegation/collaboration state, and business-level oracle are held fixed; only a machine representation changes from BusinessRules to an FMU.

\begin{table}[H]
\centering
\caption{Worked Semantic Delta for a BusinessRules $\rightarrow$ FMU migration. The overall comparison is retained with caveats because new temporal/observational/failure semantics are introduced while the business authority boundary remains stable.}
\label{tab:deltaworked}
\scriptsize
\begin{tabularx}{\textwidth}{p{2.0cm}p{2.8cm}p{3.0cm}Y}
\toprule
Delta dimension & BusinessRules condition & FMU condition & Classification / rationale\\
\midrule
Time & discrete business step & solver time, step size, synchronization point & with-caveat: timing semantics changed\\
Observation & exact abstract machine state & sampled variables with tolerance & with-caveat: observation uncertainty introduced\\
Action binding & symbolic schedule command & schedule parameters bound to FMU inputs & preserving if binding is versioned/tested\\
Authority & World/contract guard before change & same guard before FMU dispatch & preserving\\
Reversibility & exact digital reset & solver-state reset under declared initialization & with-caveat unless reset equivalence is demonstrated\\
Safety/failure & rule violation only & thermal/dynamic/solver failure modes & with-caveat: new failure classes may change outcome\\
Hidden variables & none beyond rule parameters & FMU parameters/initial state/solver tolerance & with-caveat; all must be frozen in Manifest\\
\bottomrule
\end{tabularx}
\end{table}

For this migration the overall classification is \textbf{comparison-with-caveat}: the same business hypothesis may be tested, but changes must not be attributed to the AUT without reporting these deltas.

A Semantic Delta becomes \emph{comparison-breaking} when a changed substrate alters the construct being tested itself---for example, when the new environment changes who holds authority, removes a mandatory counterparty, or changes the success/fulfillment predicate. The same rule applies to participant migration: replacing a Real Human with an Agent Proxy is a participant-realization change and any induced timing, authority, observability, or behavior-distribution change must be declared.

Table~\ref{tab:participantdelta} gives the symmetric participant-side worked example. The same Real Human $\rightarrow$ Agent Proxy migration can be acceptable for an authority-protocol hypothesis yet comparison-breaking for a user-preference or acceptance hypothesis. Semantic Delta is therefore interpreted relative to the declared hypothesis, not to the infrastructure in the abstract.

\begin{table}[t]
\centering
\caption{Worked participant Semantic Delta for Real Human $\rightarrow$ Agent Proxy.}
\label{tab:participantdelta}
\scriptsize
\begin{tabularx}{\textwidth}{p{2.3cm}YYp{3.0cm}}
\toprule
Delta dimension & Real Human condition & Agent Proxy condition & Classification / rationale\\
\midrule
Decision latency & naturally distributed human response time & program/model response time & with-caveat; interaction timing changes\\
Preference origin & actual participant preference/decision & generated or calibration-conditioned behavior & breaking for preference-realism claims; caveat for protocol-only claims\\
Authority semantics & named principal/role decision rights & same role enacted through proxy artifact & preserving only if decision rights and grant semantics are unchanged\\
Reproducibility & lower; participant behavior may vary across repeats & higher under frozen proxy artifact/seed & with-caveat; repeatability changes\\
Behavioral variance & natural and participant-specific & model/rule/calibration dependent & with-caveat; distribution changes\\
Consent / data basis & direct participation and consent record & proxy calibration/use policy and provenance & with-caveat; grounding source changes\\
\bottomrule
\end{tabularx}
\end{table}

\section{Agentic Living Lab Reference Test Harness Architecture}

\begin{figure}[t]
\centering
\includegraphics[width=0.98\linewidth]{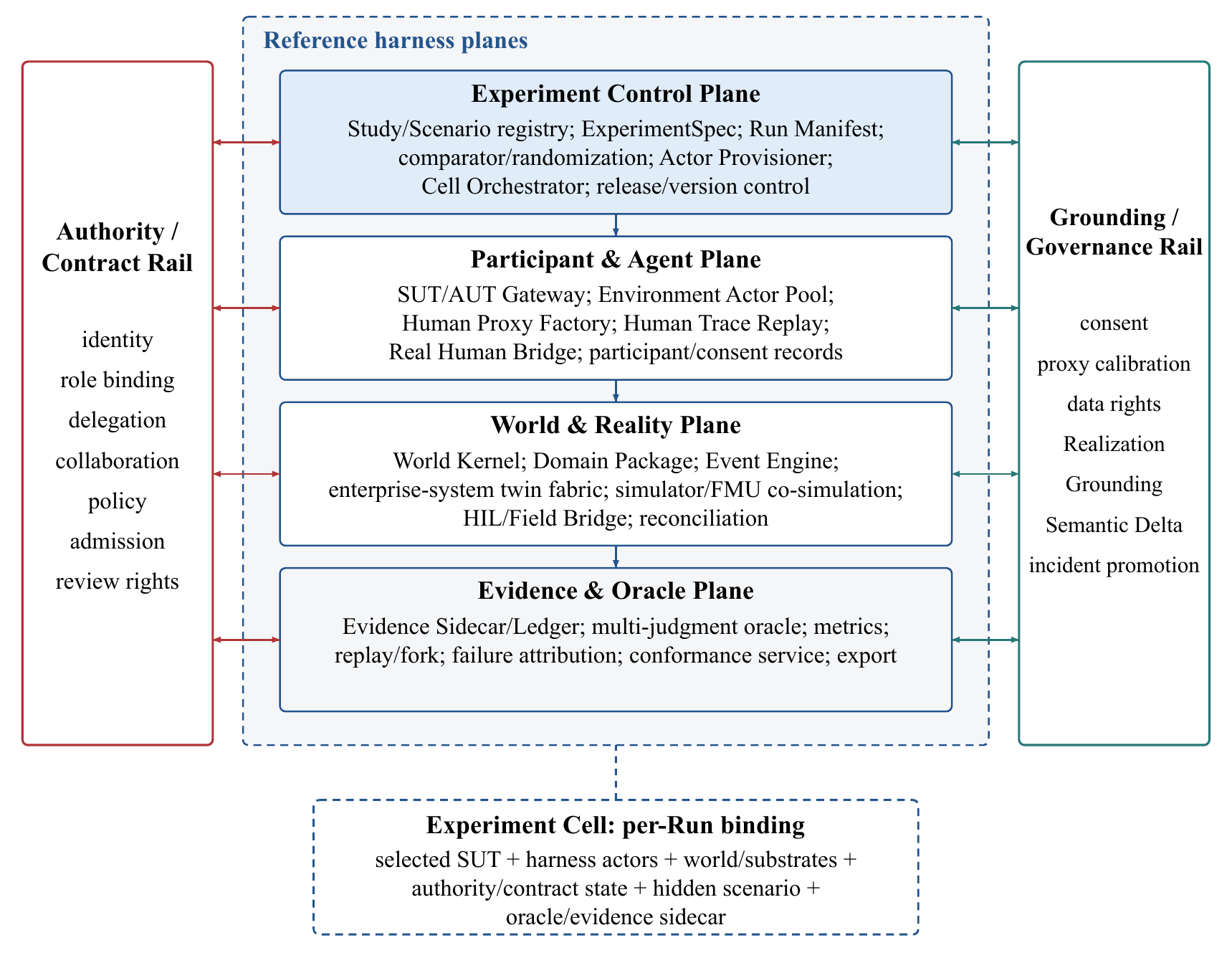}
\caption{Reference test harness architecture. Human participation is explicit through the Real Human Bridge and human-role realizations; high-throughput roles can be dynamically provisioned as proxies. Authority/contract and grounding/governance are cross-cutting controls rather than optional platform add-ons.}
\label{fig:architecture}
\end{figure}

The testing methodology is implemented by a reference \emph{test harness architecture}. The architecture is richer than a benchmark loop but deliberately thinner than a production-agent platform: it specifies the services required to configure actors and worlds, execute controlled service episodes, admit consequential actions, collect evidence, and compute multi-judgment verdicts. Figure~\ref{fig:architecture} shows four core planes and two cross-cutting rails.

\begin{figure}[t]
\centering
\includegraphics[width=0.95\linewidth]{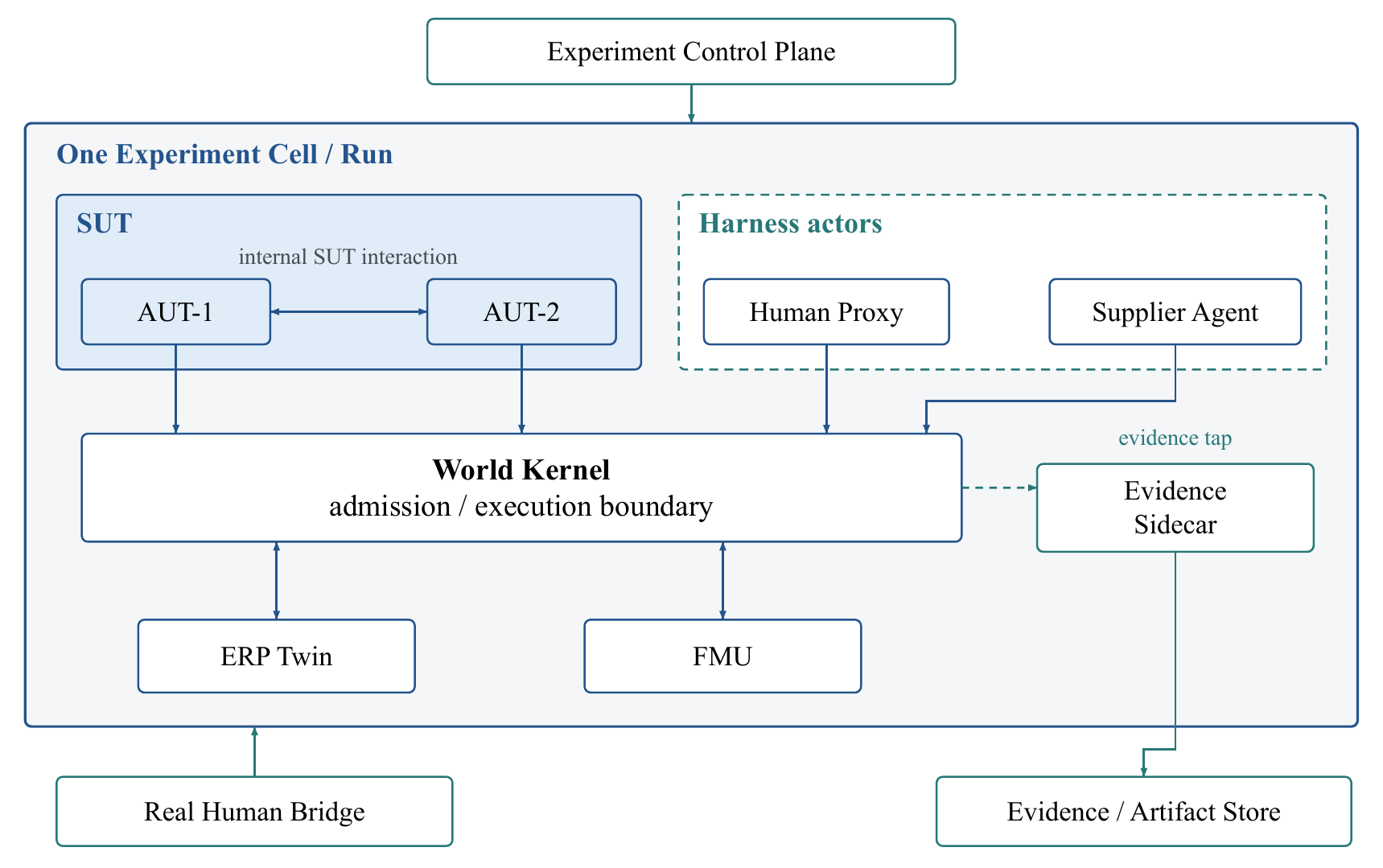}
\caption{Run deployment view. Multiple AUTs can be internal to one SUT; environment agents and human proxies are provisioned by the harness. The same Cell binds the world/substrate and Evidence Sidecar while the Control Plane, Real Human Bridge, and artifact store remain external services.}
\label{fig:deployment}
\end{figure}

\subsection{Experiment Control Plane}

The Control Plane turns a research question into versioned, repeated tests. It maintains a Study registry, Domain/Scenario packages, ExperimentSpecs, comparators and falsification conditions, and Run Manifests. A \term{Run Planner} expands experimental factors into concrete Runs; an \term{Actor Provisioner} binds SUT artifacts and harness actors; a \term{Cell Orchestrator} creates isolated Experiment Cells; and a release registry records the Lab version and reusable research assets. This plane is where randomization, matched comparisons, proxy-sensitivity designs, and cross-spectrum migrations are declared rather than improvised after results are observed.

The control object hierarchy is
\begin{equation}
LivingLab\supset Study\supset ExperimentRun,
\label{eq:hierarchy}
\end{equation}
with
\begin{equation}
ProjectSpec\rightarrow ExperimentSpec\rightarrow RunManifest.
\label{eq:specchain}
\end{equation}
ProjectSpec defines problem/governance scope. ExperimentSpec defines SUT boundary, actor realizations, variables, comparators, hidden disturbances, oracle dimensions, grounding, and falsification conditions. Run Manifest freezes the concrete artifacts and versions used by one execution attempt.

\subsection{Participant \& Agent Plane}

This plane is the key architectural addition required by the human-in-the-loop interpretation of a Living Lab.

\begin{itemize}
  \item \term{SUT/AUT Gateway}: adapts the tested system to Observation/ActionIntent interfaces without assuming its internal cognitive architecture.
  \item \term{Environment Actor Pool}: hosts counterparty agents, conventional services, scripted actors, or independently built systems that influence the episode but are not under test.
  \item \term{Human Proxy Factory}: dynamically provisions user/approver/operator/reviewer proxies from declared generators, models, prompts, behavior policies, and calibration profiles. The resulting actors are harness components and are version-pinned like other test dependencies.
  \item \term{Human Trace Replay}: re-enacts recorded human decisions/interactions under their declared context and provenance.
  \item \term{Real Human Bridge}: exposes role-scoped interfaces for delegation, approval, rejection, revocation, preference amendment, takeover, observation, review, dispute, and closure acceptance. It links decisions to consent/identity and Evidence Records.
\end{itemize}

The plane allows a Study to mix real humans and proxies. For example, 10,000 repeated travel Runs may use dynamically generated traveler proxies, while a smaller calibration/acceptance phase uses real travelers. A production Study may use an operator proxy for routine scheduling Runs but require a real safety reviewer for HIL closure. The architecture therefore avoids both extremes: it neither requires a person to sit in every simulation loop nor treats people as invisible parameters.
When a SUT contains multiple AUTs, their mutual A2A messages, negotiation, shared planning, and internal service calls are classified as internal SUT behavior. The harness may capture them for diagnostics, but they are not environment-provided stimuli. The same message becomes an environment interaction only when the Study places the counterparty outside the SUT boundary. This boundary is frozen in the Participant Realization Profile and is therefore auditable across re-runs.

\subsection{World \& Reality Plane}

The World \& Reality Plane hosts the state and consequences against which the SUT acts. The \worldkernel{} coordinates recorded business state, event ordering, invariants, admission inputs, and factual reconciliation; it is not an omniscient reality oracle. Domain Packages define entities, services, events, rules, and test fixtures. A Scenario/Event Engine injects versioned visible and hidden disturbances. Enterprise-twin adapters provide ERP/MES/WMS-like behavior or shadow access. A co-simulation fabric hosts mechanistic models and FMUs. HIL/Field Bridges expose bounded physical observation/action interfaces.

Every consequential interaction crosses a controlled boundary. The SUT does not receive direct authoritative write access merely because it owns a tool credential. World realization is object-level: one Run can bind an ERP shadow, MES twin, FMU machine, physical sensor, and software counterparty simultaneously.

\subsection{Evidence \& Oracle Plane}

Testing agentic service systems requires more than tracing model calls. The Evidence \& Oracle Plane collects canonical Evidence Records at consequential admission/effect/reconciliation boundaries and derives test judgments from them. Core components are:
\begin{itemize}
  \item an \term{Evidence Sidecar/Ledger} that stores actor binding, ActionIntent, admission decision, world/action/fact state, contract references, timestamps, provenance, and review links;
  \item a \term{Multi-Judgment Oracle} that evaluates the vector in Eq.~\eqref{eq:reportvector} rather than only task success;
  \item replay and counterfactual-fork services for digital states and model-based reconstructions;
  \item a failure-attribution service that separates Agent, Authority, Contract, WorldModel, Environment, Substrate, Evaluator, Governance, and HumanProxy causes;
  \item a conformance service that validates canonical interchange records and deterministic fixtures.
\end{itemize}

The oracle can combine automatic checks with authorized human review. A human adjudicator's decision is itself evidence with a role, policy, and version; it is not silently substituted for ground truth. If two Labs use different reconciliation or adjudication policies, that difference is part of the experimental configuration.

\subsection{Cross-cutting rails: authority/contract and grounding/governance}

The \term{Authority/Contract Rail} spans all planes. It binds identities to principals and roles, carries versioned delegation and collaboration contracts, compiles applicable guards, and defines who can approve, revoke, take over, review, dispute, or accept closure. A Human Proxy can only exercise the rights attached to the role it realizes; being generated by the Lab does not grant extra authority.

The \term{Grounding/Governance Rail} records participation and consent, human-proxy calibration, grounding in data, processes, parameters, and outcomes, Realization Profiles, Semantic Delta, incident review, and feedback assetization. These controls are what keep ``Living'' and ``realistic'' from being rhetorical labels.

\subsection{Experiment Cell and external interfaces}

A \term{Experiment Cell} is one isolated Run. It binds the SUT, environment actors, human-role realizations, world/substrates, current authority/contract state, hidden scenario configuration, evidence sidecar, and oracle. Isolation prevents one Run from contaminating another and lets different test conditions reuse the same Lab release.

The architecture can be implemented through different protocols. A minimal logical interface surface is summarized in Table~\ref{tab:interfaces}.

\begin{table}[t]
\centering
\caption{Logical interface surface of the reference test harness. Concrete transports are replaceable.}
\label{tab:interfaces}
\scriptsize
\begin{tabularx}{\textwidth}{p{3.0cm}p{3.8cm}Y}
\toprule
Interface family & Representative operation & Semantic requirement\\
\midrule
Control & create Study and ExperimentSpec; instantiate Run; stop, retry, or fork & freeze versions, factors, actors, hidden config, comparator, retention\\
SUT/actor & get Observation; submit ActionIntent; emit/receive role event & identify SUT vs harness actor; preserve principal/role/carrier and episode context\\
Human bridge & delegate, approve, revoke, takeover, review, dispute, accept & role-scoped decision + consent/identity + evidence link\\
World/reality & commit business action; read twin; step FMU; bounded HIL or field action & authority gate before consequence; declared substrate and reconciliation policy\\
Evidence/oracle & append Evidence Record; query verdict; export bundle; replay or fork & canonical provenance, mandatory violations visible, reviewer decisions attributable\\
\bottomrule
\end{tabularx}
\end{table}

The architecture is implementation-independent. OpenAPI, MCP, A2A, CloudEvents, workflow engines, policy systems, agent runtimes, and twin stacks can all be used underneath. Compatibility is judged at the normative episode/interchange/conformance boundaries, not by product names.

\section{Runtime Test Semantics and Conformance Protocol}
\label{sec:runtime}

\subsection{From participant events and ActionIntent to evidence-coupled verdicts}

The runtime accepts a common \term{EpisodeEvent} envelope with event classes such as \texttt{OBSERVATION}, \texttt{ACTION\_INTENT}, \texttt{PARTICIPANT\_DECISION}, \texttt{ENVIRONMENT\_EVENT}, and \texttt{RECONCILIATION/REVIEW}. A SUT/AUT receives Observations and may submit an ActionIntent. Harness actors---including real humans or human proxies---issue participant decisions such as delegate, approve, reject, revoke, amend preferences, take over, dispute, review, or accept closure. Those decisions can change delegation, collaboration, policy/review state, or the next Observation. They are canonical episode events with evidence, but they are not scored as AUT actions unless the Study explicitly includes that actor in the SUT. The ancillary package publishes a participant-event Schema and a human-revocation conformance fixture so this boundary is machine-inspectable rather than only diagrammatic.

\begin{figure}[t]
\centering
\includegraphics[width=0.98\linewidth]{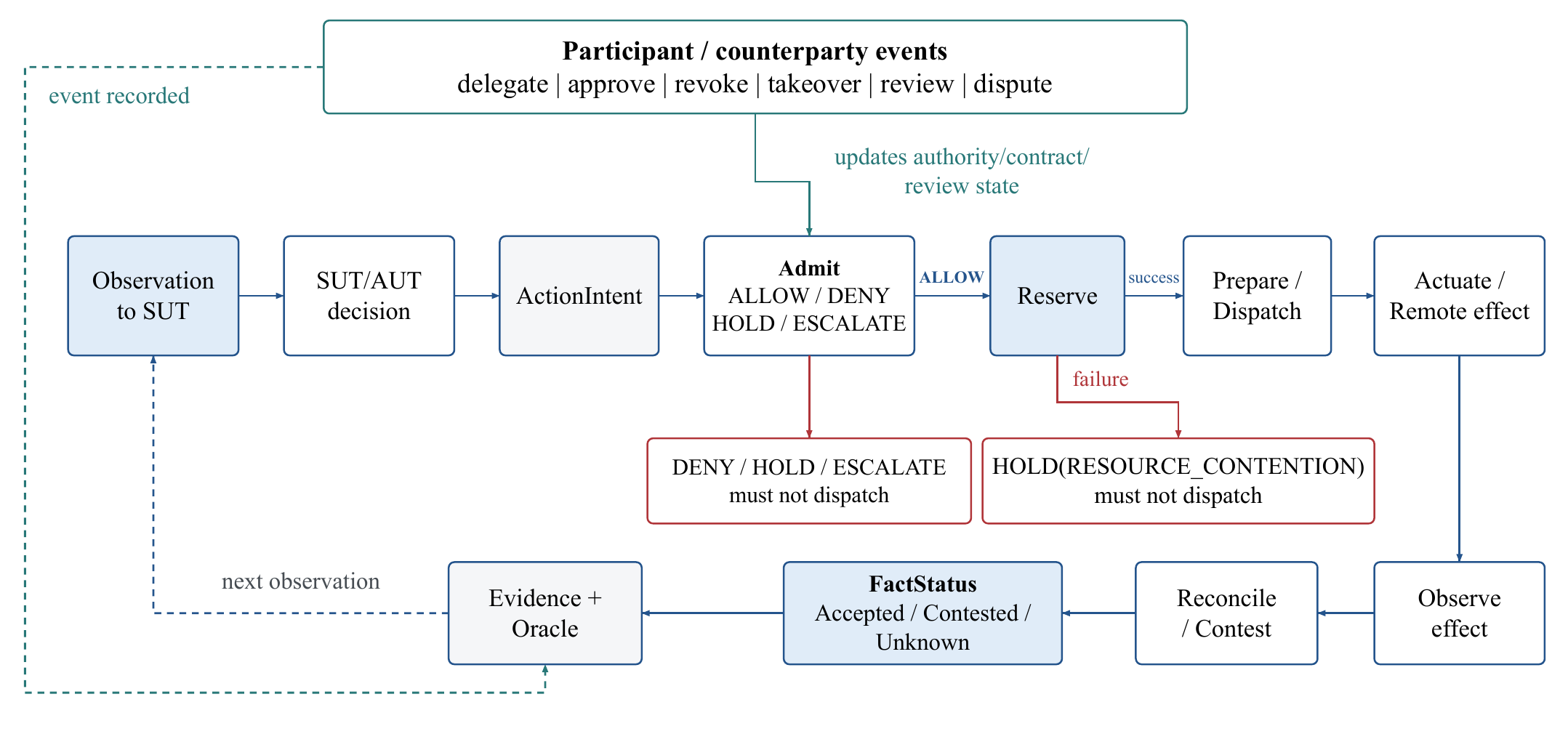}
\caption{Runtime boundary for testing. AUT intents and harness role events are distinct. Human/proxy decisions can modify authority, contracts, or review state and are recorded as episode evidence; only admitted consequential intents may proceed to effects.}
\label{fig:runtime}
\end{figure}

Figure~\ref{fig:runtime} shows the two paths. The most important invariant is that \emph{proposal, human/institutional decision, permission, remote/physical effect, factual acceptance, and fulfillment remain distinct states}.

The SUT first receives a policy-filtered observation
\begin{equation}
O_t=Projection(S_t,principal,role,disclosure).
\end{equation}
It may reason arbitrarily and emit an \actionintent{} containing at least an actor/role binding, target, operation, parameters, a correlation identifier, and where applicable an expected world version and idempotency key. Human/proxy role events use their own canonical role-decision records or domain events and can update delegation, approval, review, or closure state before a fresh AUT admission.

Admission is evaluated using the fail-closed algebra of Section~\ref{sec:minikernel}. The Lab does not publish a second Boolean formula that competes with that algebra. The resolver is total under the declared guard set: a hard \texttt{FAIL} dominates all other outcomes and yields \texttt{DENY}; absent a hard failure, an \texttt{UNKNOWN} that requires new authority or institutional judgment dominates a merely waitable unknown and yields \texttt{ESCALATE}; a remaining waitable \texttt{UNKNOWN} yields \texttt{HOLD}; only an all-\texttt{PASS} applicable guard set yields \texttt{ALLOW}. Equivalently, the decision priority is \texttt{DENY $\succ$ ESCALATE $\succ$ HOLD $\succ$ ALLOW}. A field is not made \texttt{PASS} by being irrelevant: the versioned Domain Package must declare it outside the applicable guard set for that operation.

Guard evaluation is deterministic with respect to the version-pinned world snapshot, policy/contract versions, and referenced evidence supplied to Admit. A guard must not perform hidden live substrate I/O. New live information first becomes an Observation (with provenance/version context) and only then participates in a fresh admission decision.

A versioned world must also define concurrency semantics. If an ActionIntent names an \code{expected_world_version} that is stale, the reference behavior is to prevent blind retry, record \code{HOLD(STALE_VIEW)}, re-observe, and perform a fresh admission. Shared consumable resources such as money, inventory, or capacity require atomic reservation, serializable transactions, leases/fencing, or an explicitly equivalent mechanism. Two locally permitted intents must not silently double-spend the same global resource. If a reservation fails after the guards have otherwise resolved to \texttt{ALLOW}, the action is not dispatched: the reference behavior is \code{HOLD(RESOURCE_CONTENTION)} followed by re-observation and fresh admission. \texttt{ALLOW} therefore means admissible subject to successful pre-dispatch reservation, not an irrevocable right to emit the side effect.

\subsection{Action lifecycle, factual authority, and reconciliation}

Remote and physical actions cannot be modeled as atomic database writes. We use the action lifecycle
\begin{equation}
\begin{aligned}
Proposed&\rightarrow Admitted\rightarrow Prepared\rightarrow Dispatched\\
&\rightarrow\{Acknowledged,Unknown\}\rightarrow Reconciled.
\end{aligned}
\label{eq:actionlifecycle}
\end{equation}
with terminal/review states such as RejectedBeforeDispatch, Contested, Compensated, or Closed as appropriate. The distinction matters under delayed acknowledgments, revocation races, duplicate retries, and physical uncertainty.

This lifecycle motivates a separation between two kinds of authority:

\begin{itemize}
  \item \term{Normative authority} answers what an actor is permitted to do. It is determined by identity, current delegation, policy, contract state, prerequisites, and invariants.
  \item \term{Factual authority} answers what the research authority currently accepts as the best evidence-backed factual record. In a software-only world, a transaction commit may serve both as effect and factual record. In an HIL or physical field, a command does not prove that the intended physical effect occurred; the Lab must reconcile controller acknowledgments, sensors, source systems, human confirmation, or other evidence.
\end{itemize}

The fact lifecycle therefore includes
\begin{equation}
\begin{aligned}
FactStatus\in\{&Proposed,Authorized,Dispatched,Observed,\\
&Reconciled,Contested,Unknown,Accepted\}.
\end{aligned}
\label{eq:factstatus}
\end{equation}
These labels are not a single mandatory linear state machine; they distinguish important epistemic boundaries. ActionState and FactStatus are orthogonal: for example, an action can be \code{Acknowledged} while the resulting factual claim remains \code{Contested} or \code{Unknown}.

\term{Reconciled} has a deliberately modest meaning: it is the best factual record currently accepted by the research authority under a declared reconciliation policy and the evidence available at that time. It is \emph{not} a claim of metaphysical or physically absolute truth. A sensor may drift, a source system may later correct an entry, and independent parties may dispute the observation. New evidence is appended as a new reconciliation/review event rather than rewriting the historical Evidence Record.

To keep reconciliation from becoming an uninspectable human oracle, every study that reconciles external or contested observations must pin a versioned \term{Reconciliation Policy}. At minimum the policy declares admissible source classes, any source precedence, the conflict rule, timeout/unknown handling, authorized review authority, and the acceptance condition. A simple policy may use source precedence but \emph{contest on unresolved conflict}; another domain may require an authorized human or institutional reviewer. The protocol does not prescribe one universal sensor-fusion algorithm, because that would be domain-invalid, but two Labs cannot claim matched factual semantics while silently using different reconciliation policies. The Evidence Record therefore carries the reconciliation-policy reference whenever factual reconciliation affects the episode judgment. A reconciliation-policy change is itself a Semantic Delta. If it can change the factual predicate used by the target hypothesis, the comparison is classified at least \emph{with-caveat}, and is \emph{comparison-breaking} when the accepted-fact criterion changes the construct being measured.

Similarly, \texttt{Accepted} must name its subject. An Evidence Record with \code{admission_decision=DENY} can legitimately use \code{fact_status=Accepted} if \code{fact_subject=admission_rejection}; this means that ``the action was rejected'' is an accepted fact, not that the proposed action occurred.

The time model separates authoritative event order from other clocks. Each authoritative event records at least a run sequence number and may additionally record domain time, substrate time, wall time, and source. The run sequence supports deterministic replay of the Lab record; it does not pretend to reconstruct an unknowable total order of all physical causes. A late observation is appended and reconciled rather than back-dated into the immutable history.

\subsection{Delegation and collaboration guards}

Delegation and collaboration address different questions. A \term{delegation contract} answers: \emph{what bounded discretion may this role exercise for this principal?} A \term{collaboration contract} answers: \emph{what do independently accountable parties owe one another, what state may they share, and how are fulfillment, breach, amendment, dispute, and closure determined?} Social commitments and commitment-based service interaction provide established foundations for the latter~\cite{yolum2002commitment,singh2009commitment}.

For testing, delegation is represented directly as a versioned record rather than as a mathematical tuple. Table~\ref{tab:guardmap} shows the minimal fields that compile into runtime guards. Additional domain fields are permitted, but a Study claiming conformance must expose the fields that determine authority scope, prohibitions, validity / revocation, resource/budget constraints, evidence readiness, and escalation.

\begin{table}[t]
\centering
\caption{Illustrative compilation from delegation fields to runtime guards.}
\label{tab:guardmap}
\scriptsize
\begin{tabularx}{\textwidth}{p{2.2cm}p{2.1cm}Yp{2.8cm}}
\toprule
Field & Guard role & Typical mandatory condition & Example reason code\\
\midrule
Authority scope & permission/scope & consequential action lies within current role scope & \code{OUT_SCOPE}\\
Prohibitions & prohibition & no explicit prohibition applies & \code{PROHIBITED}\\
Validity / revocation & validity / revocation & grant active and authority chain current & \code{STALE_GRANT}\\
Budget / resource & budget / resource & shared reservation succeeds & \shortstack{\texttt{BUDGET}\\\texttt{EXCEEDED}}\\
Risk & risk & declared risk predicate passes or is escalated & \code{RISK_LIMIT}\\
Evidence requirements & evidence readiness & mandatory evidence present or waitable & \shortstack{\texttt{EVIDENCE}\\\texttt{UNKNOWN}}\\
Escalation & escalation & authorized target/route exists when needed & \shortstack{\texttt{REAUTH}\\\texttt{REQUIRED}}\\
Goals / obligations / context / review & readiness / closure / review & domain-specific acceptance, obligation, provenance, dispute conditions & domain-specific\\
\bottomrule
\end{tabularx}
\end{table}

The two contract types do not cancel each other automatically. Revoking a delegation changes prospective authority. It does not erase a collaboration obligation already created under an authorized protocol. That obligation must be amended, discharged, transferred, remedied, disputed, or retained according to the collaboration agreement. Conversely, the existence of a supplier commitment does not authorize a customer's software agent to exercise rights inside the supplier's organization.

A minimal delegation lifecycle is
\begin{equation}
Draft\rightarrow Active\rightarrow\{Suspended,Revoked,Expired\},
\end{equation}
while a compact collaboration lifecycle is
\begin{equation}
\begin{aligned}
Negotiating&\rightarrow Committed\rightarrow InFulfillment\\
&\rightarrow\{Fulfilled,Breached,Cancelled,Disputed\}\rightarrow Closed.
\end{aligned}
\end{equation}
The two are linked by episode events, not by a single shared status.

\subsection{HOLD, ESCALATE, contested closure, and reference algorithm}

The operational semantics of non-ALLOW outcomes are important for independent implementations.

\paragraph{HOLD.} A HOLD decision records the unresolved mandatory condition, a nonempty wake condition and deadline, and an on-expiry disposition: DENY, ESCALATE, or UNRESOLVED in the reference interchange profile. UNRESOLVED preserves a non-dispatched unresolved outcome; it never authorizes execution. When a wake condition fires or policy requests a new evaluation, the old decision is never resumed blindly: all guards are evaluated again against the then-current world and authority state. Thus a grant revoked while the action was held produces DENY under the total priority order, even if the original missing evidence has since arrived.

\paragraph{ESCALATE.} ESCALATE names an authorized principal, reviewer, or higher role capable of making the missing decision. The original action remains blocked. A later approval does not mutate the old decision; it creates or amends authority and triggers a fresh admission.

\paragraph{Contested/unknown closure.} If a mandatory fact or obligation remains Contested or Unknown, the episode cannot be labeled Fulfilled. The Lab may record a terminal status such as Closed-with-dispute, Cancelled, Breached, or Unresolved if the governing protocol permits it, but residual obligations and the unresolved claim remain explicit.

Algorithm~\ref{alg:intent} summarizes the ordering constraints. It is pseudocode, not a requirement to use any particular programming language or policy engine.

\begin{algorithm}[tp]
\caption{Reference processing of an EpisodeEvent}
\label{alg:intent}
\begin{algorithmic}[1]
\Require episode event $e$, pinned world/policy/contract state, participant profile, substrate $R$
\If{$e$ is a participant decision}
  \State validate actor/role decision right and current authority
  \State append canonical participant event + evidence
  \State update delegation/collaboration/review state if valid
  \State wake affected held/escalated intents, but require fresh admission against the new state
  \State \Return updated episode state
\EndIf
\If{$e$ is an environment/observation event}
  \State append observation with provenance; update only the recorded observable state
  \State wake affected held intents; require fresh admission if a consequential action is reconsidered
  \State \Return updated episode state
\EndIf
\State assert $e$ contains an ActionIntent $I$
\State $g\gets EvaluateGuards(I,current\ state)$
\State $d\gets ResolveFailClosed(g)$
\State record admission decision and reasons
\If{$d=DENY$} \State \Return rejected-before-dispatch \EndIf
\If{$d=HOLD$} \State register wake/deadline/expiry policy; \Return held-before-dispatch \EndIf
\If{$d=ESCALATE$} \State issue authorized review request; \Return escalated-before-dispatch \EndIf
\State $r\gets TryAcquireReservations(I,concurrencyPolicy)$
\If{$r=FAILED$}
  \State record \code{HOLD(RESOURCE_CONTENTION)}; re-observe and require fresh admission
  \State \Return held-before-dispatch
\EndIf
\State $a\gets PrepareAndDispatch(I,R)$
\State $o\gets ObserveEffect(a,R)$
\State $f\gets ApplyVersionedReconciliationPolicy(o)$
\State append Evidence Record linking intent, admission, action/effect, reconciliation, and versions
\State \Return updated recorded state, fact status, and evidence links
\end{algorithmic}
\end{algorithm}
\FloatBarrier

\subsection{Reality substrates, rollback, and physical compensation}

Software engineers are accustomed to snapshots, resets, and branches. Physical systems are not Git repositories. A real robot collision, shipment, or human decision cannot be erased by restoring a database snapshot. The protocol therefore distinguishes
\begin{equation}
DigitalRollback\neq PhysicalCompensation.
\end{equation}
A software world may restore a prior state for a new experimental branch. A real field event can only be followed by additional physical/business actions that compensate, repair, or otherwise respond to the consequence.

The same distinction applies to counterfactual forks. A digital experiment can fork $S_k$ into multiple branches. A physical study usually creates a digital reconstruction of the observed state and performs the counterfactual in a twin, FMU, or simulator. The resulting branch is evidence about a model-based counterfactual, not a claim that physical history was rerun.

Industrial bindings can reuse established standards. FMI/FMU supplies model-exchange and co-simulation interfaces~\cite{fmi2024}; OPC UA and Asset Administration Shell specifications supply industrial information and asset representations~\cite{opcua2026,idta2025}. The contribution here is not the standards themselves but the way a versioned model identity, variable binding, solver settings, state/parameter snapshot, authority context, and evidence are frozen inside the experiment protocol.

\subsection{Experiment Cell runtime, scenario semantics, and evidence boundary}

An \term{Experiment Cell} is an isolated execution unit that hosts one Run and enforces trust boundaries. A minimal deployment separates:

\begin{itemize}
  \item an authority core containing World Kernel/time/policy-contract guards;
  \item a hidden/evidence zone containing evaluator secrets, holdout conditions, Evidence Sidecar, and protected scenario state;
  \item a controlled runtime zone containing experimental agents, role agents, enterprise twins, or FMU runners;
  \item an external/untrusted zone for third-party agents, tool proxies, or adapters.
\end{itemize}

The architecture requires that possession of a tool credential alone confer no direct write authority over authoritative state. This requirement assumes complete mediation of consequential writes, isolated credentials and adapters, and an uncompromised authority and evidence boundary. It is not a demonstrated guarantee against prompt injection or malicious code: an attack may exploit an implementation defect or produce an admissible but harmful intent. Security assurance therefore requires separate implementation and adversarial validation.

A scenario is likewise more than a prompt:
\begin{equation}
Scenario=StateModel+EventGraph+Roles+Rules+Tools+Evaluator+DisclosurePolicy.
\label{eq:scenario}
\end{equation}
Hidden disturbances---a future flight cancellation, a capacity shock, a machine fault---can be version-pinned in the scenario without being disclosed to the agent. The evaluator may know the ground truth that the agent is not authorized to observe.

Every consequential boundary yields evidence. A run-level \evidencebundle{} may contain
\begin{equation}
\begin{aligned}
EB=\{&Manifest,Observations,Intents,AdmissionDecisions,Actions,Facts,\\
&Contracts,Versions,Hashes,Metrics,Reviews\}.
\end{aligned}
\label{eq:evidencebundle}
\end{equation}
This is deliberately stronger than runtime logs. Runtime logs can be implementation-specific and incomplete; an authoritative Evidence Record is a canonical research artifact associated with a boundary decision/effect; derived evaluation is a later interpretation of those records.

OpenAPI, MCP, A2A, and CloudEvents can be used at interface boundaries~\cite{openapi2025,mcp2026,a2a2026,cloudevents2022}, but transport success is not world authority. HTTP 200, MCP tool success, A2A task completion, or CloudEvent delivery can be evidence of communication or invocation; they do not by themselves establish an authorized business transition, a fulfilled obligation, or a physically verified effect.

\section{Test Workflow and Constructed Experiment Designs}

\subsection{Study lifecycle}

\begin{figure}[t]
\centering
\includegraphics[width=0.98\linewidth]{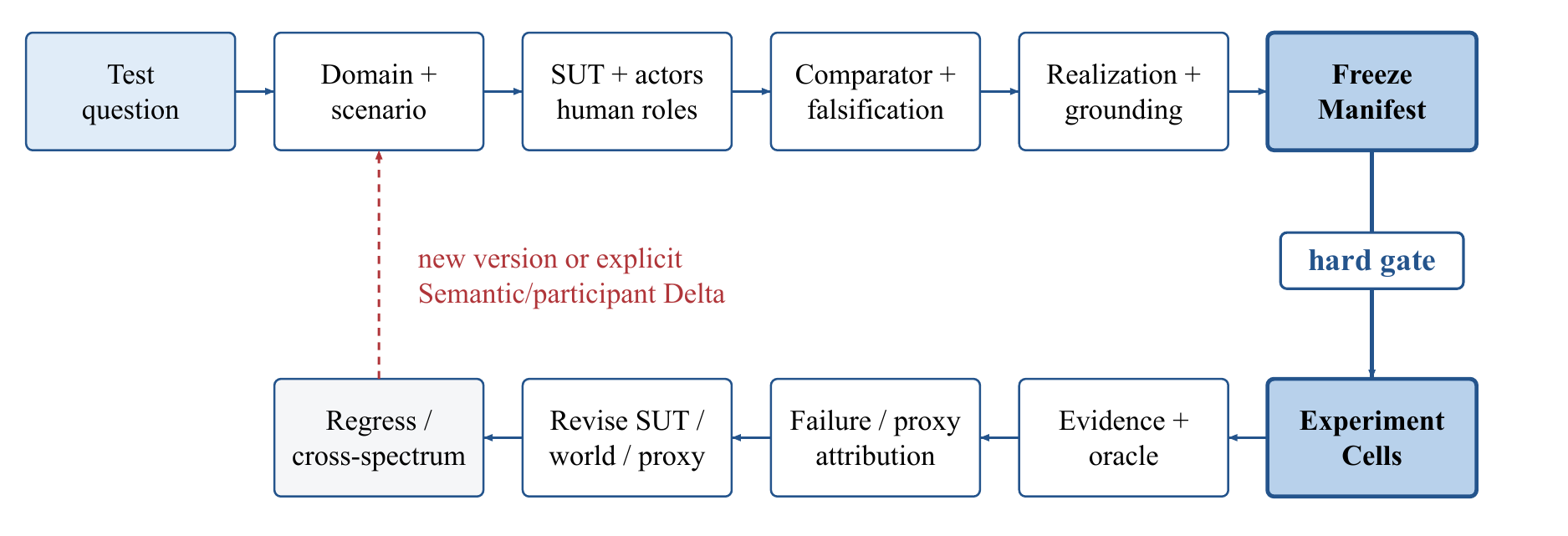}
\caption{Testing and engineering workflow. Actor realization is configured explicitly so that AUT effects can be separated from environment-agent or human-proxy effects.}
\label{fig:workflow}
\end{figure}

The methodology is intended for both evaluators and agent/service-system developers. Figure~\ref{fig:workflow} shows the standard loop.

A Study follows seven operational phases.

\paragraph{1. Test question and baseline.} Declare the claim, a strong comparator, controlled variables, and a falsification condition. Conventional automation, an adaptive workflow with policy/compensation/audit, a task-oriented agent, or a mechanism ablation can all be valid comparators.

\paragraph{2. Domain and scenario.} Select a versioned Domain Package defining entities, states, services, events, rules, policies, evaluators, fixtures, and migration rules. A Scenario binds the initial world, hidden disturbances, disclosure policy, and termination/closure conditions.

\paragraph{3. SUT and actor realization.} Define the SUT/AUT boundary. Provision environment agents/services and map each human role to Real Human, Human Trace Replay, Rule Surrogate, or Agent Proxy. Freeze proxy generators/artifacts and declare which real stakeholders ground or review the Study.

\paragraph{4. Comparator and experimental controls.} Select the mechanism/agent/environment factor to vary while freezing the factors that could otherwise confound the comparison. If simulated human behavior matters, predeclare the proxy-sensitivity comparison or the human evidence used to justify the proxy.

\paragraph{5. Reality and grounding configuration.} Select object-level Realization and Grounding Profiles. Record the physical/digital substrate, participant realization, and any Semantic Delta relative to a comparator or prior Run.

\paragraph{6. Freeze and execute.} The Run Manifest pins the SUT artifact, harness actors, policies, contracts, world snapshot, seed, substrate bindings, evaluator, concurrency/reconciliation semantics, and evidence/retention policy. Experiment Cells execute the runtime cycle and produce canonical evidence.

\paragraph{7. Evaluate, attribute, revise.} The oracle derives the multi-judgment vector; failures are attributed to SUT, authority, contract, world, environment, substrate, evaluator, governance, or human-proxy causes; the SUT/world/proxy/hypothesis is revised; and matched Runs are repeated.

\subsection{Replay and counterfactual fork}

For a deterministic recorded world, replay should reconstruct the same authoritative state from the same initial snapshot and accepted event sequence:
\begin{equation}
Replay(M,E)\Rightarrow hash(S_n')=hash(S_n).
\label{eq:replay}
\end{equation}
This does not require the LLM to regenerate identical tokens. Replay reconstructs the authoritative history, including accepted/rejected/uncertain transitions and their evidence links.

A \term{Counterfactual Fork} starts from a frozen digital state $S_k$ and changes one declared factor. Examples include agent version, contract-guard mode, one hidden disturbance, or a policy parameter. A fork supports controlled comparison, but it does not by itself prove causal identification; the study must still declare randomization/control assumptions, repeats, and confounders.

For physical field studies, the fork normally occurs after constructing a digital representation of the observed state. The resulting experiment is labeled accordingly in the Realization/Grounding Profiles and Semantic Delta.

\subsection{Failure attribution}

A binary failure label is too coarse for agent engineering. We use the diagnostic taxonomy
\begin{equation}
\begin{aligned}
Failure\in\{&Agent,Authority,Contract,WorldModel,\\
&Environment,Substrate,Evaluator,Governance,HumanProxy\}.
\end{aligned}
\label{eq:failuretaxonomy}
\end{equation}
The categories are not always mutually exclusive; a review can assign primary and contributing causes. Their purpose is to distinguish ``the agent chose poorly'' from ``the experiment omitted a real constraint'' or ``the source system produced an unresolved observation.''

For example, a travel agent that proposes a purchase beyond its delegation budget is primarily an Agent/Authority failure if the grant was correctly represented and the guard correctly denies it. If a valid accessibility rule is absent from the Domain Package, the failure may instead be WorldModel. If two source systems disagree on whether a physical shipment occurred, the episode may remain Environment/Substrate/Contested rather than being forced into an agent-success or agent-failure label.

\subsection{Feedback assetization}

A Living Lab differs from a static benchmark because real operation can alter future research assets. The feedback loop is
\begin{equation}
RealWorld\rightarrow Incident/Case\rightarrow Review\rightarrow VersionedAsset\rightarrow NextLabRelease.
\label{eq:assetization}
\end{equation}
An incident is not copied directly into a benchmark. It first passes source/permission checks, minimization/de-identification, expert review, causal reconstruction where possible, and ethics/safety review. Only then can it become a Scenario, Rule, Validator, Parameter Set, Golden Run, or calibration asset.
\begin{figure}[t]
\centering
\includegraphics[width=0.90\linewidth]{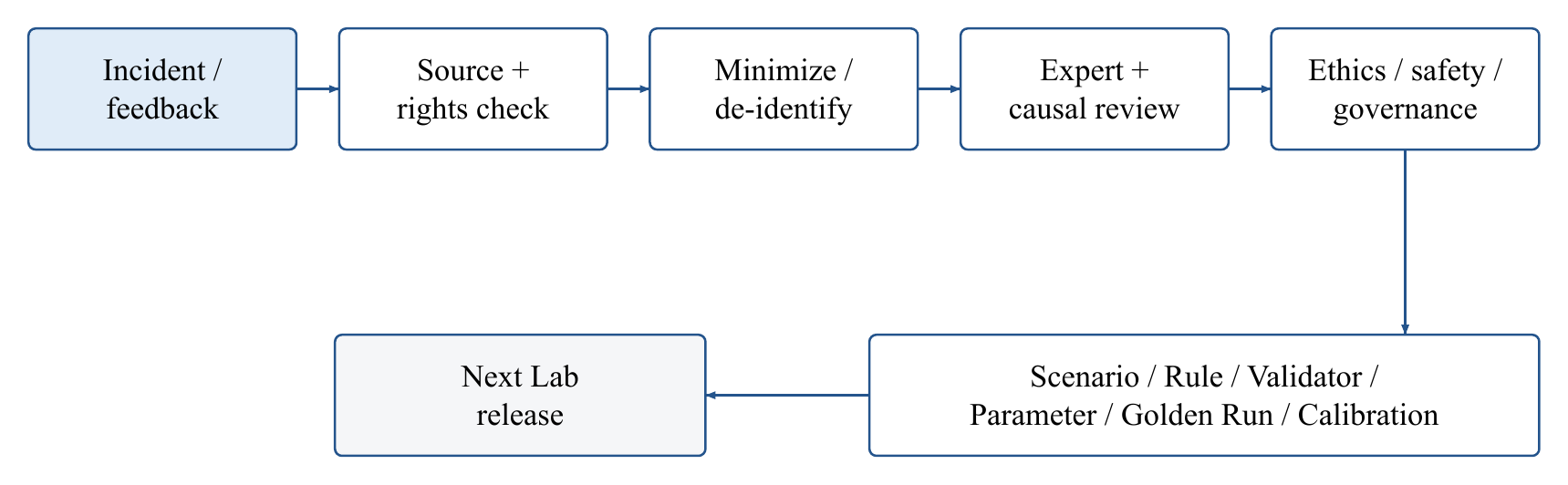}
\caption{Feedback assetization. Real incidents become reusable research assets only after provenance, rights, causal, and governance review.}
\label{fig:feedback}
\end{figure}

This loop gives ``Living'' a technical meaning without reducing it to perpetual online learning. The Lab evolves through reviewed, versioned releases rather than silently changing the world under ongoing experiments.

\subsection{Travel-service agent: from fulfillment witness to controlled ablation}

\subsubsection{Study setup}

Suppose a developer is building a \term{Travel Disruption Recovery Agent}. A traveler has a multi-leg international trip involving air travel, hotel accommodation, ground transport, insurance, and payment. The service objective is not merely to output an itinerary; it is to maintain a service episode under disruptions while respecting delegated authority and provider commitments.

The travel Domain Package contains at least the following participants and service objects:
\begin{itemize}
  \item Traveler principal and a delegated Travel Agent role;
  \item Airline, Hotel, Ground Transport, Insurance, and Payment provider roles;
  \item bookings, holds, tickets, payments, refunds, accessibility claims, and approval records;
  \item service rules for cancellation, rebooking, late arrival, refund, and evidence/closure.
\end{itemize}

The traveler issues a delegation that allows comparison and reversible holds, permits automatic rebooking only within a declared extra-cost budget, and requires renewed approval for a nonrefundable purchase above the delegated threshold. Provider commitments remain distinct: an airline's transport obligation and a hotel's accommodation obligation arise only through their accepted protocols.

The ExperimentSpec freezes a hidden disturbance such as
\begin{equation}
FlightCancelled(t=30\ \mathrm{min}),
\end{equation}
without disclosing that future event to the agent. Other hidden factors may include hotel late-arrival policy, seat availability, or a temporary payment-service delay.

\subsubsection{A constructed witness: same output, different fulfillment}

The first use of the Lab is analytical rather than statistical. Construct two traces that produce the same final itinerary and the same final purchase price.

\paragraph{Trace A: controlled progression.} The agent obtains required approval before the nonrefundable purchase, atomically reserves the shared travel budget, reconciles an old hold before rebooking, retains provider confirmation for accessibility, and closes the episode only after the authorized reviewer accepts the evidence and residual obligations are discharged.

\paragraph{Trace B: unresolved relational violation.} The agent buys the same ticket without required approval, copies task state to a replacement carrier while the old carrier retains authority, relies on an unverified accessibility claim, and marks the trip ``complete'' while an obligation/hold remains unresolved.

The two traces establish the constructed witness
\begin{equation}
Output_A=Output_B,\qquad Fulfillment_A\neq Fulfillment_B.
\label{eq:witness}
\end{equation}
This is not an empirical superiority claim. It demonstrates that, under a contract semantics in which authority and evidence constrain fulfillment, output equivalence is insufficient for fulfillment equivalence. The ancillary package binds Trace B to the interchange profile through \code{examples/travel-trace-b-illegal-closure.json}; the reference conformance runner projects its missing approval/evidence state to the same non-fulfilled closure outcome as fixture C5. This demonstrates representability and machine checking of the constructed witness, not practical superiority.

\begin{table}[t]
\centering
\caption{Constructed output-equivalent traces.}
\label{tab:travelwitness}
\scriptsize
\begin{tabularx}{\textwidth}{p{2.7cm}YY}
\toprule
Boundary & Trace A: controlled & Trace B: unresolved\\
\midrule
Purchase & required approval recorded & same purchase without required approval\\
Shared budget & atomic reservation across branches & local branch checks allow temporary over-commitment\\
Disruption recovery & old authority reconciled/revoked before replacement & task state copied while old carrier authority remains active\\
Evidence & provider accessibility confirmation retained & unverified text copied into the plan\\
Closure & authorized review; residual obligations resolved & ``trip complete'' despite unresolved obligation/hold\\
\bottomrule
\end{tabularx}
\end{table}

\subsubsection{Failure diagnosis and agent revision}

Now consider an actual agent artifact $A_1$ whose heuristic is ``after cancellation, select the earliest feasible replacement flight.'' After observing the cancellation, the agent proposes a replacement that exceeds the delegated extra-cost limit. The admission layer returns \texttt{DENY} before dispatch.

A binary task report might say only that recovery failed. The Evidence Bundle instead supports the chain
\begin{equation}
FlightCancelled\rightarrow RecoveryPlan\rightarrow ChangeFlightIntent\rightarrow DelegationViolation\rightarrow DENY.
\end{equation}
The diagnosis is therefore not ``the agent cannot search flights.'' It is that authority constraints were not treated as hard planning conditions before the proposal was generated.

The developer may revise the agent to $A_2$ by incorporating active delegation and contract state into candidate generation. A further version $A_3$ may reason about downstream service dependencies, such as whether a six-hour delay threatens hotel check-in or insurance/ground-transport commitments. The Lab supports this engineering loop because the same world, hidden disturbance, and evaluator can be reused across versions.

\subsubsection{A strict single-variable ablation}

A v1--v2--v3 development story is useful for engineering, but it is not by itself a clean causal comparison. To isolate the value of a relational control, define a mechanism ablation in which the agent artifact, model, prompt, memory snapshot, tools, world snapshot, service inventory, seed, delegation, and hidden events are held fixed. Only one mechanism changes:
\begin{equation}
ContractGuard\in\{OFF,ON\}.
\end{equation}
The ExperimentSpec predeclares the falsification condition. If enabling the guard does not reduce unauthorized dispatch or unresolved obligations under matched conditions---or if its additional rejection/review burden creates greater total service loss over the intended scope---the practical claim that the control provides net value must be narrowed.

\begin{table}[t]
\centering
\caption{Travel-case comparison discipline. No performance numbers are asserted.}
\label{tab:travelablation}
\scriptsize
\begin{tabularx}{\textwidth}{p{3cm}YY}
\toprule
Comparison & Held fixed & Changed / question\\
\midrule
Contract-guard ablation & agent artifact, model/prompt/tools, world, seed, delegation, hidden events & guard OFF/ON; does the runtime prevent unauthorized or unfinished episodes?\\
Agent revision & world, seed, substrate, contracts, evaluator & $A_1$ vs $A_2$/$A_3$; does planning improve under the same relational constraints?\\
Grounding migration & agent and experiment logic as far as justified & synthetic persona vs real-user shadow; do acceptance/trust and violation patterns change?\\
\bottomrule
\end{tabularx}
\end{table}

\subsubsection{Digital grounding without physical facilities}

Travel is also a useful counterexample to the idea that a Living Lab must become physically situated. The study can migrate from synthetic travelers to real-user shadow operation while remaining entirely digital. Real travelers submit real needs, the agent produces plans and proposed intents, but the Lab does not execute purchases. User acceptance, takeover burden, consent, policy violations, and explanation quality can then be measured against governed real participation.

The resulting study has low physicality but potentially high stakeholder/process grounding. This is exactly why physicality and grounding are reported separately.

\subsection{Human-role proxy sensitivity test}

Proxy sensitivity is evaluated as a matched test, not as an informal warning. The Study freezes the AUT/SUT version, Scenario, world snapshot, contracts, environment actors, disturbances, and primary oracle measures, and changes only the realization of one human role. At minimum it uses two behaviorally distinct proxy realizations; where feasible, a human trace or smaller Real Human condition is added as a grounding anchor.

For a single AUT compared with a fixed baseline, let $\Delta m_P$ denote the difference on primary measure $m$ under proxy realization $P$. A direction change,
\begin{equation}
\operatorname{sign}(\Delta m_{P_1})\neq \operatorname{sign}(\Delta m_{P_2}),
\label{eq:proxysign}
\end{equation}
acts as a \term{proxy-sensitivity reporting trigger}, not as an inferential significance test. For multiple AUTs, any reversal of the predeclared ranking on a primary measure is likewise a mandatory \term{rank-instability trigger}. These triggers require the paper to expose the dependence instead of hiding it inside an aggregate score; they do not by themselves establish statistical significance or faithful human modeling. Empirical claims should add prespecified repetitions, uncertainty intervals or effect magnitudes, and an appropriate inferential procedure when the Study scale permits.

The Participant Realization Profile reports the proxy artifact/version and calibration asset for every condition. A triggered result is bounded to the tested participant realizations until further human grounding is obtained. This test does not establish that any proxy is a faithful model of humans; it identifies whether the reported agent-quality conclusion may depend on how the human role was instantiated.

\subsection{Industrial agent: separating substrate sensitivity from agent improvement}

\subsubsection{Study setup}

Consider a \term{Production Recovery Agent} for a manufacturer that must respond to supplier delay and a critical-machine disturbance while coordinating purchasing, production scheduling, inventory, and logistics. The episode involves multiple principals---customer, manufacturer, supplier, and logistics provider---and a mixture of enterprise information systems and physical constraints.

A first software-native run may bind
\begin{equation}
ERP=Twin,\quad MES=Twin,\quad WMS=Twin,\quad Machine=BusinessRules.
\end{equation}
The machine model may expose only availability, capacity, processing time, and a few state transitions. The agent can therefore produce a schedule that is business-feasible in this digital world.

\subsubsection{Substrate-sensitivity experiment}

The first controlled question is not whether a new agent is better. It is whether the conclusion is sensitive to the environment representation. Keep the agent, scenario, contracts, evaluator, seed, and organizational state fixed, but replace
\begin{equation}
Machine: BusinessRules\rightarrow FMU.
\end{equation}
The FMU is version-pinned together with model hash, solver, step size, tolerance, initial state, parameters, and variable/unit bindings. FMI provides the model-exchange/co-simulation standard; the experimental contribution is the versioned binding and evidence context rather than FMI itself~\cite{fmi2024}.

Suppose the same schedule that satisfies business rules now violates a thermal constraint in the mechanistic model. The observation demonstrates the logical possibility
\begin{equation}
BusinessFeasible\not\Rightarrow MechanisticallyFeasible.
\label{eq:businessphysical}
\end{equation}
Again, this is a constructed example, not a measured field result. Its purpose is to show how substrate sensitivity can invalidate a conclusion even when the agent artifact has not changed.

The migration must record a Semantic Delta: continuous time is introduced; additional state becomes observable; new safety/feasibility constraints exist; rollback semantics change; and the evaluator now receives mechanistic evidence. Without this declaration, the study would incorrectly describe the two runs as ``the same experiment'' except for an implementation detail.

\subsubsection{Agent-improvement experiment}

Only after the environment is fixed at the FMU realization should the developer test a mechanism-aware agent revision. Keep the FMU, world, disturbance, contracts, and evaluator fixed, and compare
\begin{equation}
A_1\rightarrow A_2,
\end{equation}
where $A_2$ evaluates candidate schedules against the mechanistic model before issuing the consequential scheduling intent. This comparison addresses an agent-design question under a common substrate rather than mixing environment and agent changes.

\begin{table}[t]
\centering
\caption{Industrial-case separation of substrate and agent variables.}
\label{tab:industrialdesign}
\scriptsize
\begin{tabularx}{\textwidth}{p{3.2cm}YY}
\toprule
Experiment & Held fixed & Changed / question\\
\midrule
Substrate sensitivity & Agent $A_1$, scenario, contracts, evaluator, seed & Business-rule machine $\rightarrow$ FMU; does the prior conclusion survive added mechanistic semantics?\\
Agent improvement & FMU and all experiment conditions & Agent $A_1\rightarrow A_2$; does mechanism-aware planning improve under the same substrate?\\
Shadow migration & Agent and declared core episode identity & Twin $\rightarrow$ real MES/asset shadow; what new operational assumptions or failure modes appear?\\
HIL migration & Agent and declared experiment controls & FMU/local simulation $\rightarrow$ HIL; do timing/control constraints change the effect?\\
\bottomrule
\end{tabularx}
\end{table}

\subsubsection{Shadow feedback and Lab self-correction}

A later study may bind a real MES and operational data in read-only shadow mode. The agent observes live state but all consequential intents remain simulation-only. Real operation can then expose assumptions absent from the Lab: temporary maintenance, shift changes, human overrides, communication delays, production insertions, or unmodeled resource interactions.

The Evidence Bundle must help distinguish at least three situations:
\begin{enumerate}
  \item the agent failed under a correctly represented constraint;
  \item the Lab's world model omitted or misrepresented a real constraint;
  \item the real environment produced ambiguous/contested evidence that prevents a definitive judgment.
\end{enumerate}
This distinction is central to an experimental infrastructure. A Lab that automatically attributes every mismatch to the agent becomes a biased evaluator of its own assumptions.

Reviewed incidents can be converted into new regression scenarios, parameter sets, validators, or calibration assets through the feedback-assetization process. Subsequent agent versions are then tested against the newly versioned Lab release.

\subsubsection{HIL and controlled physical validation}

If the research question genuinely depends on real controller timing, communication delays, sensor noise, or hardware execution, one local component can move from FMU to HIL while the rest of the study remains digital. The question becomes: \emph{does the agent effect observed under the FMU realization persist under the newly declared timing, actuation, safety, and observability semantics?}

The Lab therefore supports staged evidence without requiring a full production line for every research question. Increasing physicality is justified by the claim being tested, not by a universal maturity narrative.

\section{Evaluation, Conformance, and Implementation Status}
\label{sec:conformance}

\subsection{Minimum reporting vector and measurement discipline}

\label{sec:metrics}
The methodology intentionally avoids reducing agentic-service quality to one scalar. It instead requires a reporting vector whose components have explicit working definitions. For a Study with $N$ evaluated episodes and $N_a$ consequential action attempts made by SUT/AUT actors under test, we recommend at least the following measures when applicable:

\begin{align}
FR &= \frac{N_{fulfilled\ episodes}}{N}, \label{eq:fr}\\
UDR &= \frac{N_{unauthorized\ dispatched\ actions}}{N_a}, \label{eq:udr}\\
EC &= \frac{N_{accepted\ mandatory\ evidence\ items}}{N_{required\ mandatory\ evidence\ items}}, \label{eq:ec}\\
HIR_{req} &= \frac{N_{episodes\ with\ prespecified\ mandatory\ human\ decisions}}{N}, \label{eq:hirreq}\\
HIR_{obs} &= \frac{N_{episodes\ with\ observed\ human\ intervention}}{N}, \label{eq:hirobs}\\
DER &= \frac{N_{AUT\text{-}triggered\ nonmandatory\ escalations}}{N}. \label{eq:der}
\end{align}
$HIR_{req}$ is fixed by the Scenario/contract/evaluator before the Run; $HIR_{obs}$ records what actually happened; and $DER$ exposes discretionary workload that the SUT pushes back to people. This prevents a conservative agent that escalates everything from appearing ``safe'' without paying a visible human-work cost. For human-involved studies, total/median active human review time $T_{review}$ and approval latency should accompany these rates. Attempted but successfully denied unauthorized actions are reported separately from UDR because they indicate risky proposals without an escaped side effect. Evidence coverage also reports counts of \texttt{Contested} and \texttt{Unknown}; an unresolved mandatory item cannot be averaged away by high coverage elsewhere.

Commitment/recovery measures are domain-specific but must define their numerator, denominator, observation window, and residual-obligation rule before the test. Stakeholder value is represented by a declared \term{Value Ledger} rather than a universal score. A minimal ledger may record accepted benefit $B$, direct service cost $C_d$, recovery/remediation cost $C_r$, human-attention cost $C_h$, realized residual loss $L$, and declared externalities $X$. When these quantities are legitimately commensurable, a Study may additionally report
\begin{equation}
NV = B-C_d-C_r-C_h-L-X,
\label{eq:netvalue}
\end{equation}
but the methodology does not require all value dimensions to be monetized or collapsed into one scalar. The Study must state whose benefit/cost is represented and the observation window over which value is assessed.

A concrete harness implementation may itself affect the tested system. When performance matters, the report should therefore include harness-overhead measures such as admission latency, evidence serialization/write latency, synchronization delay, and storage amplification. This paper reports no such latency or storage numbers and makes no claim that the harness overhead is negligible.

\begin{table}[t]
\centering
\caption{Minimum reporting vector. A paper may add measures but should not hide mandatory violations in a composite score.}
\label{tab:reporting}
\scriptsize
\begin{tabularx}{\textwidth}{p{2.8cm}Yp{4.3cm}}
\toprule
Dimension & Observable measure & Required qualification\\
\midrule
Outcome / obligation & FR; accepted goal achievement; residual obligations & separate Fulfilled, Failed, Cancelled, Breached, Disputed, Unresolved\\
Authority / constraints & UDR; attempted violations; budget/prohibition violations & distinguish blocked attempts from dispatched side effects\\
Commitment / continuity & recovery time/rate; duplicate effects; surviving obligations after substitution/revocation & state observation window and commitment-status rule\\
Evidence / contestability & EC; number of Contested/Unknown mandatory items; review resolution & log completeness alone does not establish factual truth\\
Human work & $HIR_{req}$; $HIR_{obs}$; DER; $T_{review}$; approval latency; takeover count & distinguish mandatory human decisions from AUT-triggered discretionary escalation\\
Stakeholder value & declared Value Ledger; optional $NV$ only when dimensions are commensurable & identify stakeholder, observation window, and non-monetized dimensions/externalities\\
Participant realization & human-role mode; proxy/trace artifact/version; calibration asset; proxy-sensitivity trigger & do not generalize one surrogate population to real humans without grounding\\
Reality / grounding & Realization Profile; Grounding Profile; Semantic Delta & do not infer external validity from physicality alone\\
Harness overhead & admission/evidence latency; synchronization delay; storage amplification & report when implementation performance can affect the conclusion; no universal ``negligible'' assumption\\
\bottomrule
\end{tabularx}
\end{table}

Repeated runs should report variation, tail loss, contract coverage, reviewer disagreement where applicable, and test-contamination controls. Studies should include conditions where conventional automation is expected to be sufficient; otherwise the benchmark embeds the preferred conclusion.

\emph{The hypotheses H1--H6 are not claims that this methodology or reference architecture has already proven; they are a minimal set of falsifiable questions that future studies using the Lab should answer, and the measurement definitions above specify how their primary outcomes must be made observable.}

\subsection{H1--H6: falsifiable questions, not established results}

\paragraph{H1: Does explicit relational state improve substitution and recovery?}
When a carrier, provider, or plan changes, does a runtime that maintains versioned grants, commitments, and evidence produce fewer unauthorized actions, unresolved obligations, and incorrect closures than a matched runtime that tracks only task state? A strong adaptive workflow with policy, compensation, and audit is a required comparator. A null result or greater total loss after control overhead would narrow the claim.

\paragraph{H2: Do dual-contract controls produce measurable net value?}
Does separating delegation authority from collaboration obligations reduce authority laundering, invalid commitments, or recovery ambiguity relative to a matched control? The evaluation must include false rejections, review cost, and human attention. A guard that blocks everything is not a successful result.

\paragraph{H3: Does a persistent world expose problems hidden by task reset?}
Compared with independent task-reset episodes, does persistent relationship/world state reveal delayed obligations, resource encumbrance, stale authority, correlated failures, or residual risk that materially affect service judgments? If not, the persistence mechanism may be unnecessary for that task class.

\paragraph{H4: Does the Evidence Bundle add diagnostic value beyond runtime logs?}
Under matched execution traces, can the canonical Evidence Records improve independent reconstruction, authority/closure judgment, failure attribution, or dispute review relative to ordinary task/tool logs? The constructed travel witness demonstrates why the distinction can matter, but comparative benefit remains empirical.

\paragraph{H5: Are measured effects stable across Reality Substrates?}
If a mechanism or agent version appears beneficial in a software realization, does the direction or magnitude of the effect persist under a declared migration to calibrated twin, shadow, FMU/HIL, or controlled physical conditions? A sign change after substrate migration is not a failed paper; it identifies the boundary of the original claim.

\paragraph{H6: Do participant realization, world realization, and grounding explain different aspects of external validity?}
When studies vary human-role realization independently from world Realization and Grounding Profiles, do these disclosures explain different kinds of error, ranking instability, or generalization? If proxy choice does not materially affect conclusions, or if simpler declarations explain the relevant outcomes just as well, the reporting scheme should be simplified.

\subsection{Normative reference-interchange schemas and conformance artifacts}

This paper makes a concrete interoperability commitment. The source package distributed with the preprint includes the following normative reference-interchange Schemas. Paths below are relative to the ancillary-artifact root rather than to the TeX source directory:

\begin{itemize}
  \item \code{schemas/run-manifest.schema.json}
  \item \code{schemas/participant-profile.schema.json}
  \item \code{schemas/evidence-record.schema.json}
  \item \code{schemas/action-intent.schema.json}
  \item \code{schemas/admission-decision.schema.json}
  \item \code{schemas/conformance-case.schema.json}
  \item \code{schemas/reconciliation-policy.schema.json}
  \item \code{schemas/participant-event.schema.json}
\end{itemize}

\term{Normative status.} The JSON Schema files are normative for the paper's reference-interchange profile. An implementation may use different internal objects, field names, databases, or languages, but an implementation claiming conformance must be able to export a deterministic canonical record with semantics equivalent to the defined profile. Internal naming freedom does not permit loss of the SUT/AUT boundary, participant-role realization, principal/role binding, current grant/contract versions, admission result, world version, action/fact state, realization/grounding declarations, or required evidence links.

JSON Schema can express structure, types, required fields, enumerations, and local constraints. It cannot by itself express all temporal semantics such as ``revocation preserves incurred obligations'' or ``Contested mandatory evidence blocks Fulfilled closure.'' Consequently, conformance is jointly defined by (i) the Schemas, (ii) the runtime transition rules in this paper, and (iii) deterministic input--expected-output fixtures.

The source package includes the following machine-readable fixtures:
\begin{itemize}
  \item \code{fixtures/C1_stale_grant/case.json}
  \item \code{fixtures/C2a_waitable_evidence/case.json}
  \item \code{fixtures/C2b_reauthorization/case.json}
  \item \code{fixtures/C4_carrier_replacement/case.json}
  \item \code{fixtures/C5_closure_evidence_missing/case.json}
  \item \code{fixtures/C6_uncertain_remote_effect/case.json}
  \item \code{fixtures/C8_human_revocation/case.json}
\end{itemize}
C3/C7 remain prose conformance oracles in this version and are explicitly not represented as released machine-readable fixtures.

The ancillary package also includes \path{tools/reference_conformance_runner.py} and tests under \path{tests/}. The runner consumes C1/C2a/C2b/C4/C5/C6/C8, applies their limited reference rules, emits reference-interchange records where applicable, validates them against the Schemas, and compares the declared scalar and list-valued oracle projections. Unsupported admission and successful-closure cases raise explicit errors rather than defaulting to ALLOW or FULFILLED. Fixture annotations named \code{forbidden_side_effects} and \code{required_evidence} specify obligations for a future integrated harness; they are reported as unevaluated, not counted as passed checks. The runner neither dispatches actions nor observes a live world and therefore cannot verify the absence of actual side effects. Evidence records retain supplied identities, relationship references, and revocation causation; missing fixture context uses documented synthetic defaults. C6 assumes prior admission as an input precondition rather than recomputing it. Boundary regression tests check priority, unsupported inputs, decision-specific Schema constraints, and evidence linkage. These are executable checks of a selected normative subset, not a World Kernel, a security proof, or empirical agent evaluation.

\begin{table}[t]
\centering
\caption{Released deterministic conformance fixtures.}
\label{tab:fixtures}
\scriptsize
\begin{tabularx}{\textwidth}{p{1.2cm}p{4.2cm}p{2.2cm}Y}
\toprule
Case & Input condition & Expected decision/status & Forbidden side effects / required interpretation\\
\midrule
C1 & stale/revoked delegation for consequential action & DENY; rejected before dispatch & no dispatch, mutation, or reservation; retain denial evidence\\
C2a & mandatory evidence unknown but explicitly waitable & HOLD & no dispatch; retain wake condition, deadline, on-expiry policy\\
C2b & current grant insufficient; authorized re-approval route exists & ESCALATE & original action blocked; retain target, grant version, reason\\
C4 & carrier replaced after old grant is revoked while an incurred commitment survives & DENY until fresh binding/grant; commitment retained & copying task state must not transfer authority or erase obligation state\\
C5 & task complete but mandatory closure evidence missing & NOT\_FULFILLED & do not mark episode fulfilled or erase residual obligations\\
C6 & remote effect dispatched; acknowledgment/effect remains uncertain & UNKNOWN / NOT\_FULFILLED & no fabricated success and no blind redispatch without idempotency/reconciliation\\
C8 & authorized human/principal revokes current delegation before a later consequential intent & participant revocation + DENY & revocation recorded as episode event; later action must use fresh authority\\
\bottomrule
\end{tabularx}
\end{table}

A conforming implementation is not required to produce byte-identical internal traces. It must produce semantically equivalent canonical records and the same expected decisions/statuses for the declared fixtures under the stated assumptions.

\subsection{Minimum empirical validation design required for stronger claims}

The present paper does not report agent-quality results. To move from a specified methodology to empirical evidence, a subsequent study should execute at least one end-to-end episode corpus rather than merely adding more schemas. A minimal travel-domain design can compare two or more runnable SUTs (for example, a deterministic workflow baseline and one or more agent implementations) under the same world snapshots, contracts, disturbances, and participant realizations. It should report ordinary task success alongside FR, UDR, EC, $HIR_{req}$/$HIR_{obs}$/DER, $T_{review}$, a declared Value Ledger, and failure attribution, and should preserve the Evidence Records that explain any task-success/fulfillment disagreement.

A second matched design can hold the SUT fixed and vary either the human-role realization (proxy sensitivity) or the Reality Substrate (BusinessRules versus FMU). These are not promises of positive findings: a null difference would narrow the need for the corresponding methodology in that task class. This section specifies the evidence needed to test the method; it is not a schedule or a claim that such experiments have already been run.

\subsection{Implementation status}

Table~\ref{tab:status} states the evidence boundary at submission time. This table is intentionally conservative.

\begin{table}[H]
\centering
\caption{Implementation/evidence status of the present paper and released artifacts.}
\label{tab:status}
\scriptsize
\begin{tabularx}{\textwidth}{>{\raggedright\arraybackslash}p{3.0cm}ZZZZ}
\toprule
Artifact/construct & Conceptual definition & Constructed fixture/example & Executable/normative specification & Running prototype / independent validation\\
\midrule
Episode/relationship experimental object & yes & yes & partial profile & no / no\\
Admission algebra & yes & yes & normative rules + reference runner subset & small runner / no independent validation\\
Delegation/collaboration guard semantics & yes & yes & partial profile & no / no\\
World Kernel/reconciliation semantics & yes & yes & reference algorithm & no / no\\
Run Manifest & yes & example & normative JSON Schema & no / no\\
Participant Realization Profile & yes & travel/proxy examples & normative JSON Schema & schema-validated examples / no independent validation\\
Evidence Record & yes & example & normative JSON Schema & no / no\\
ActionIntent/\allowbreak AdmissionDecision & yes & fixture inputs/outputs & normative JSON Schema & runner emits/validates released admission cases; no independent validation\\
Conformance cases C1/\allowbreak C2a/\allowbreak C2b/\allowbreak C4/\allowbreak C5/\allowbreak C6/\allowbreak C8 & yes & yes & machine-readable fixtures + pytest runner & reference runner passes / no independent validation\\
Realization/\allowbreak Grounding/\allowbreak Semantic Delta & yes & case profiles incl. participant realization & schema fields / reporting rules & no / no\\
Reality Substrate contract / Reality Binding & yes & object-level Manifest example & normative operation semantics + JSON Schema & no adapters / no independent validation\\
FMU/HIL/field adapters & design only & industrial illustration & interface expectations only & no / no\\
\bottomrule
\end{tabularx}
\end{table}

The absence of a running Living Lab prototype is not hidden. This version specifies the canonical Reality Substrate operation semantics and validates object-level Reality Bindings as Manifest data, but it does not provide an adapter SDK or claim working FMU/HIL/field integrations. It adds only a small reference conformance runner for the released fixture subset; it does not implement persistent world execution, remote reconciliation, a production Human Proxy Factory/Real Human Bridge, FMU/HIL adapters, or stakeholder governance. The next engineering target is an independently implemented minimal kernel that reproduces the same admission, fact/closure, binding, and evidence judgments without sharing the runner code. Comparative benefits under H1--H6 require later measurements.

\section{Discussion and Conclusion}

\subsection{Why a Living Lab if the paper is about testing?}

The methodological repositioning does not make the Living Lab concept decorative. It clarifies its role. A generic test harness can generate tasks, run agents, and compute scores. An Agentic Living Lab additionally keeps the test program connected to the socio-technical system in which the agentic service will matter: stakeholders can co-define goals and failure criteria; real principals can delegate/revoke/approve; human roles can be calibrated or replaced by proxies whose validity is disclosed; disputed outcomes can be reviewed by identified authorities; and real incidents can be governed into future test assets.

The repeated Run need not contain a real human every time. What matters is that human roles and rights are explicit and that the test program has a path from real stakeholders to proxy calibration, acceptance, challenge, and feedback. This is analogous to the physical--digital continuum: physical presence is one realization, not the definition of reality or of Living-Lab completeness.

\subsection{Why not simply call this Agent testing?}

The paper does not attempt to define a universal test theory for all autonomous agents. Its scope is narrower: \term{agentic service systems} in which software actors represent identifiable principals, participate in service relationships, invoke or compose services, mutate persistent state, and can incur obligations or physical consequences. This service-computing boundary is what makes delegation, collaboration commitments, fulfillment, and stakeholder value part of the oracle rather than optional application metadata.

A generic tool-use benchmark can therefore remain perfectly useful without implementing this protocol. The method becomes relevant when a test claim depends on authorization, cross-party obligations, human decisions, persistent/contested effects, or migration across digital and physical realizations.

\subsection{Reproducibility, human realism, and external validity}

Digital tests provide reset, fork, fixed seeds, controlled hidden events, and inexpensive repetition. Real-human and physical conditions provide authentic timing, preferences, workarounds, noise, institutions, and consequences. Neither axis monotonically dominates the other.

Human proxies make the tradeoff especially visible. They provide repeatability and scale but can distort cooperation, language style, risk perception, decision timing, and escalation behavior. The methodology therefore treats proxy choice as a test variable. A result obtained against one Agent Proxy is a statement about that declared participant realization unless human traces, alternative proxies, or real-participant conditions support broader generalization.

Similarly, Realization Profile, Grounding Profile, and Semantic Delta scope claims about world realism. ``Real hardware,'' ``real user,'' ``real data,'' and ``real business acceptance'' are distinct sources of grounding.

\subsection{Boundaries to adjacent systems}

The reference infrastructure is not an AgentOps platform. AgentOps can provide deployment, tracing, monitoring, cost, and lifecycle functions; the Lab adds SUT/comparator declarations, harness actors, human-role realization, hidden disturbances, service relationships, authority/obligation state, multi-judgment oracles, proxy sensitivity, substrate disclosure, falsification, replay/fork, and governed feedback assets~\cite{dong2024agentops}.

It is not a replacement for stateful agent benchmarks. $\tau$-bench, $\tau^2$-Bench, STATE-Bench, CRMArena, and related environments already demonstrate stateful interaction, active/simulated users, policies, and mutable worlds~\cite{yao2025taubench,barres2025tau2,microsoft2026statebench,huang2025crmarena}. They can host parts or all of the method if the required episode/participant/authority/evidence semantics are made explicit. The claim is portability of the testing contract, not exclusive capability.

It is not a renamed Digital Twin System. Digital-twin frameworks increasingly include agentic actuation, authority, and evidence~\cite{dtc2026framework}. The Lab adds test-control objects, SUT/harness separation, participant realizations, comparators, randomized/hidden disturbances, proxy sensitivity, conformance fixtures, and episode-level service judgments. Twins, FMUs, and HIL are Reality Substrates inside the method.

\subsection{Limitations and next empirical step}

This version is a testing-method and reference-architecture report. The travel and industrial designs are constructed; the ancillary Schemas, fixtures, and small conformance runner demonstrate machine-readable semantics for a subset but do not validate the practical superiority of the method or the completeness of the platform.

Four limitations remain central. First, independent implementations must still demonstrate that they reach semantically equivalent admission, reconciliation, closure, and evidence judgments. Second, Human Proxy validity is empirical and role-dependent; simulator choice can change test results. Third, Reconciliation Policy makes factual adjudication inspectable but does not solve sensor fusion, epistemology, or legal fact-finding. Fourth, Semantic Delta discloses changes across substrates but does not prove that two Runs are causally comparable.

The next empirical milestone can therefore be much smaller than a full Living Lab deployment. A travel-domain study can run several existing or independently implemented agents under the same episode protocol, compare ordinary task success with the multi-judgment oracle, and repeat the same AUTs across a rule user, two agent-based human proxies, human-trace replay, and a small real-human shadow/calibration sample. A second, equally compact industrial study can hold the AUT fixed while replacing a business-rule machine model with an FMU, then hold the FMU fixed while comparing an agent revision. These studies would directly test whether the methodology exposes failures/ranking changes that task-only evaluation misses, without requiring a production HIL installation.

\subsection{Conclusion}

The central contribution of this paper is a way to \emph{test} agentic service systems, not a claim that a complete new platform has already been deployed. The method asks three questions: what episode is being tested, who enacts the relevant software and human roles, and under what reality/grounding is the episode executed. The Agentic Living Lab is the reference infrastructure that makes those questions executable.

The resulting operational chain is
\begin{equation}
Observation\rightarrow ActionIntent\rightarrow Admission\rightarrow Execute/Reconcile\rightarrow Evidence\rightarrow Oracle,
\end{equation}
and the engineering loop is
\begin{equation}
\begin{aligned}
TestQuestion&\rightarrow ConfigureActors/World\rightarrow Run\rightarrow Evidence\rightarrow Verdict\\
&\rightarrow Attribution\rightarrow Revision\rightarrow MatchedReTest\rightarrow GroundingMigration.
\end{aligned}
\end{equation}

A useful output is not automatically an authorized action; a completed Run is not automatically a fulfilled service episode; an active Human Proxy is not automatically a faithful human; a protocol message is not automatically a business fact; and a physical facility is not automatically stronger evidence than a well-grounded digital study. By forcing these distinctions into the test configuration and oracle, the methodology aims to make agentic-service evaluation more falsifiable, diagnosable, and portable across the environments in which service consequences matter.

\appendix

\section{Normative Run Manifest and Evidence Record Profile Summary}
\label{app:schemas}

The complete normative Schemas are distributed with the source package in the \code{schemas/} directory. This appendix summarizes the canonical fields to make the paper readable without reproducing several pages of JSON.

\subsection*{Run Manifest}

A canonical Run Manifest contains at least:
\begin{itemize}
  \item \code{run_id}, \code{study_id}, and \code{lab_release};
  \item world binding: Domain Package, world snapshot, scenario version, and seed;
  \item SUT/AUT artifact information: artifact identifier and, where applicable, model, prompt hash, memory snapshot, and tool profile;
  \item Participant Realization Profile: SUT/AUT membership, environment actors, human roles, their Real Human/Trace/Rule/Agent-Proxy realizations, decision rights, calibration references, and optional Real Human Bridge;
  \item relationship binding: delegation version and collaboration-contract versions;
  \item policy/evidence profiles, including a versioned Reconciliation Policy when external/contested observations affect factual judgment;
  \item concurrency semantics: world-version policy, reservation semantics, remote idempotency, and unknown-effect handling;
  \item time model: authoritative order, domain time, wall time, and optional substrate time;
  \item Realization Profile, Grounding Profile, and optional Semantic Delta;
  \item object-level Reality Bindings that freeze binding/adapter versions, substrate type, Capability Descriptor, endpoint or artifact, I/O/time binding, authority/reconciliation/safety policies, and any Semantic Delta reference;
  \item experiment/evaluator configuration and retention policy.
\end{itemize}

\subsection*{Reality Substrate and object-level Reality Binding}

The normative Run Manifest Schema represents each non-participant world/service/asset realization as a versioned \code{reality_bindings} entry. The entry contains an object identifier, binding identifier/version, substrate type, adapter identity/version, endpoint or artifact, a complete Boolean Capability Descriptor for the canonical operations, optional constraints, I/O and time-binding references, authority/reconciliation/safety policy references, and an optional Semantic Delta reference. A Capability Descriptor states what an adapter claims to support; conformance to the operational semantics in Table~\ref{tab:substratecontract} remains an implementation property. Human-role carriers remain in the Participant Realization Profile and are not encoded as Reality Substrates.

\subsection*{Participant Realization Profile}

The normative source package includes \code{schemas/participant-profile.schema.json}. It identifies the SUT and AUT set, environment actors, human-role realizations, role-scoped decision rights, proxy generator/calibration references, and an optional Real Human Bridge carrying consent/identity policy. When a proxy is calibrated from human data, the profile records the calibration asset reference/hash, population/sampling description, calibration procedure, permitted use, access/release status, and privacy constraints. The profile is a test-dependency record: an Agent Proxy can be dynamically generated by the Lab without becoming part of the SUT.

\subsection*{Evidence Record}

A canonical Evidence Record contains at least:
\begin{itemize}
  \item run/episode identifiers and authoritative sequence number;
  \item principal/role/carrier binding and relevant grant/contract/policy versions;
  \item ActionIntent identifier and admission decision/reason codes;
  \item action lifecycle state and, if applicable, fact subject/status;
  \item before/after world versions or hashes where meaningful;
  \item observation/effect/reconciliation references and the reconciliation-policy reference when applicable;
  \item evidence/provenance links and timestamps;
  \item correlation/causation identifiers sufficient to reconstruct the episode boundary.
\end{itemize}

Implementations may use richer records. The normative claim is that the canonical meaning above must be exportable without semantic loss for fields required by the relevant conformance case.

\section{Conformance Cases and Fixture Index}
\label{app:protocol}

The source package currently releases seven deterministic machine-readable fixtures, a small reference conformance runner, and additional cases as prose oracles. The runner is intentionally smaller than a Lab kernel; it exists to make the released normative subset executable and inspectable.

\subsection*{Released fixtures}

\begin{description}[leftmargin=1cm,style=nextline]
\item[C1 stale grant.] A consequential intent under a revoked/stale delegation must yield DENY and remain rejected before dispatch. No world mutation or reservation is allowed. The denial itself is recorded as an accepted factual boundary.
\item[C2a waitable evidence.] Unknown mandatory evidence that is explicitly waitable must yield HOLD with wake condition, deadline, and on-expiry policy. Dispatch is forbidden.
\item[C2b reauthorization.] Current authority is insufficient but an explicit authorized re-approval route exists. The expected decision is ESCALATE; the original action stays blocked and a later approval triggers fresh admission.
\item[C4 carrier replacement after revocation.] The old carrier's grant has been revoked, a successor carrier is introduced without a fresh binding/grant, and a previously incurred collaboration commitment remains active. The successor action must be denied until fresh authority exists; the surviving commitment must remain visible and must not be erased by task-state copying.
\item[C5 closure evidence missing.] A completed task with missing mandatory closure evidence cannot produce Fulfilled closure. Residual obligations/evidence gaps remain visible.
\item[C6 uncertain remote effect.] A dispatched remote/physical action whose acknowledgment/effect cannot be established remains Unknown. The harness must not fabricate success or blindly redispatch when idempotency/reconciliation requirements are unresolved.
\item[C8 human revocation.] A principal/human-role participant revokes an active grant as an episode event. A later consequential ActionIntent under the revoked grant must be denied; the event remains distinct from AUT behavior.
\end{description}

\subsection*{Prose conformance oracles retained for future machine fixtures}

\begin{description}[leftmargin=1cm,style=nextline]
\item[C3 shared-resource concurrency.] Two individually permitted intents contend for one non-duplicable resource. At most one reservation may succeed under the declared concurrency policy; the other must re-admit against the new state.
\item[C7 contested mandatory fact.] A mandatory claim under contestation blocks Fulfilled closure until the applicable review process resolves or explicitly terminates the episode as non-fulfilled/disputed.
\end{description}

A complete future conformance suite should provide machine inputs/expected outputs for C3/C7 and independent implementations in at least two runtimes.

\section{Protocol and Standard Mapping}
\label{app:standards}

Table~\ref{tab:standardmap} is a reuse and integration map, not a novelty or precedence table. The listed standards and research families already provide important interface, provenance, normative, lifecycle, simulation, and twin mechanisms; this protocol specifies how selected capabilities are bound to a versioned service episode, its authority/obligation state, substrate declarations, and evidence boundaries.

\begin{table}[h]
\centering
\caption{Representative standards/technologies and the Lab integration question.}
\label{tab:standardmap}
\scriptsize
\begin{tabularx}{\textwidth}{p{3.1cm}p{4.2cm}Y}
\toprule
Concern & Representative anchor & Lab integration requirement\\
\midrule
HTTP service interfaces & OpenAPI~\cite{openapi2025} & bind operations/side effects to ActionIntent, current authority, and evidence semantics\\
Agent--tool/context access & MCP~\cite{mcp2026} & tool capability does not grant world authority; preserve principal/role context\\
Agent--agent interoperability & A2A~\cite{a2a2026} & map task/message lifecycle to service relationship, delegation, commitments, and closure\\
Events & CloudEvents~\cite{cloudevents2022} & event delivery is evidence of transport, not automatic authoritative fact\\
Human/user simulation & $\tau$-bench, $\tau^2$-Bench, SimulatorArena, RealUserSim~\cite{yao2025taubench,barres2025tau2,dou2025simulatorarena,zhu2026realusersim} & separate human-role proxy from AUT; freeze proxy artifact/calibration and report proxy sensitivity when conclusions depend on it\\
Provenance / research objects & W3C PROV, SIMPROV, RO-Crate, CWL~\cite{moreau2013prov,ruscheinski2025simprov,soilandreyes2022rocrate,crusoe2022cwl} & package replayable artifacts while linking them to consequential admission/effect/reconciliation boundaries\\
Norms / ongoing usage control & normative MAS, UCON$_{ABC}$~\cite{boella2006normative,park2004ucon} & compile applicable institutional/authorization/obligation conditions into explicit episode guards and reason codes\\
Industrial information & OPC UA / AAS~\cite{opcua2026,idta2025} & bind asset identity/state to organizational authority and evidence\\
Mechanistic simulation & FMI/FMU~\cite{fmi2024} & freeze model/solver/parameters/bindings and report substrate-induced Semantic Delta\\
Digital Twin System & DTC framework~\cite{dtc2026framework} & reuse authority/evidence/actuation concepts while adding study/comparator/replay/falsification semantics\\
Business-artifact lifecycle & GSM artifacts~\cite{hull2011gsm} & reuse explicit long-running data/lifecycle semantics while retaining principal/authority/commitment and substrate/evidence bindings\\
\bottomrule
\end{tabularx}
\end{table}
\FloatBarrier

\printbibliography

\end{document}